\documentclass[aps,prd,twocolumn,showpacks,superscriptaddress,ref,floatfix]
{revtex4-2}
\usepackage{amsmath,amssymb,marginnote}
\usepackage{graphicx,subfigure}% Include figure files
\usepackage{dcolumn}% Align table columns on decimal point
\usepackage{bm}% bold math
\usepackage[colorlinks=true, allcolors=blue]{hyperref}

\usepackage[usenames]{color}

\newcommand{\liu}[1]{\textcolor{black}{#1}}
\begin{document}
	\newcommand\barparent[1]{\overset{%
			\scriptscriptstyle(-)}{#1}}
	
	% \preprint{APS/123-QED}
	
	\title{A Systematic Study on the \liu{Resonance-like Structure} in Collisional Neutrino Flavor Instability}% Force line breaks with \\
	
	% \author{Jiabao Liu, Shoiichi Yamada, Masamichi Zaizen}
	% \collaboration{High Energy Astrophysics Group Waseda University}%\noaffiliation
	\author{Jiabao Liu}
	\affiliation{Department of Physics and Applied Physics, School of Advanced Science \& Engineering, Waseda University, Tokyo 169-8555, Japan}
	\author{Masamichi Zaizen}
	\affiliation{Faculty of Science and Engineering, Waseda University, Tokyo 169-8555, Japan}
	\author{Shoichi Yamada}
	\affiliation{Department of Physics, School of Advanced Science \& Engineering, Waseda University, Tokyo 169-8555, Japan}
	\affiliation{Research Institute for Science and Engineering, Waseda University, Tokyo 169-8555, Japan}
	\date{\today}% It is always \today, today,
	
	\begin{abstract}
		Investigations on the \liu{resonance-like phenomenon} in the collisional flavor instability (CFI) of neutrinos, which were \liu{observed in the linear phase} recently, are reported. We show that \liu{it} occurs not only for the isotropy-preserving modes as pointed out in the previous work but also for the isotropy-breaking modes and that it enhances the \liu{linear} growth rate of CFI. % by orders of magnitude 
		Employing the linear analysis and nonlinear numerical simulations in the two-flavor scheme and under the relaxation approximation for the collision term, we discuss the criterion for the \liu{resonance-like phenomenon observed in the linear phase}, its effect on the \liu{subsequent} nonlinear evolution as well as the influences of homogeneity-breaking ($k \ne 0$) perturbations as well as of anisotropy in the background on the \liu{resonance-like structure}. We will also touch the cohabitation of the \liu{resonance-like structure} with the fast flavor conversion (FFC).
		%Neutrino-matter non-forward scattering, referred to as the collision term, was thought to decohere neutrinos and damp flavor instability, but recently it was pointed out that the collision term can indeed drive flavor instability when the system is stable to other effects. It was found that this instability can be enhanced by roughly two orders of magnitudes under certain stringent fine-tuning conditions through a resonance. We show analytically for a homogeneous, isotropic, and monochromatic neutrino gas the reason for this resonance, the exact criteria, and a formula for strength of instability under various physical conditions. The monochromatic formula is found to be a good approximation to non-monochromatic neutrino gases to some extents. The resonance brings the system by two orders of magnitudes faster to saturation, whereas the saturation level of flavor coherence is approximately a constant as long as initial neutrino number densities are not changed significantly. In linear regime, we confirm that the instability in non-monochromatic neutrino gases behaves energy-independently. Collisional instability in homogeneous and isotropic background is found to be suppressed slightly by anisotropy in the background and strongly by propagating perturbation. As an extreme case to anisotropy where fast flavor instabilities are generated, the overlap between collisional resonance and fast instability can potentially explain the controversy on whether their interplay enhanced or suppresses instabilities.
	\end{abstract}
	% \begin{description}
		% Secondary publications and information retrieval purposes.
		% You may use the \texttt{description} environment to structure your abstract;
		% use the optional argument of the \verb+\item+ command to give the category of each item. 
		% \end{description}
	
	\maketitle
	
	\section{\label{sec:level1}Introduction}
	% Neutrino flavor oscillations and especially fast flavor conversion have been studied extensively owing to its fast timescale of action and strength compared to other instability schemes.
	Neutrino flavor oscillations, particularly the fast flavor conversion (FFC), have been studied extensively these days due to their more rapid growth compared to other conversion modes\,\cite{PhysRevD.46.510,Sigl1993GeneralKD,doi:10.1146/annurev.nucl.012809.104524,CHAKRABORTY2016366,doi:10.1146/annurev-nucl-102920-050505,https://doi.org/10.48550/arxiv.2207.03561}. The neutrino-flavor-lepton-number (NFLN) crossing, that is, the situation\liu{,} in which one flavor is dominant over the other in certain momentum directions while the opposite is true in other directions at least for a pair of neutrino flavors, is known to be the criterion for FFC\,\cite{PhysRevD.105.L101301}. In its application to core-collapse supernovae (CCSNe), the electron-lepton-number (ELN) crossing has been conveniently searched for in many papers\,\cite{PhysRevD.100.043004,Nagakura_2019,PhysRevD.101.023018,PhysRevD.101.043016,PhysRevResearch.2.012046,PhysRevD.101.063001,Abbar_2020,PhysRevD.103.063013,PhysRevD.104.083025,Harada_2022,https://doi.org/10.48550/arxiv.2206.01673}, since other flavor-lepton-numbers are normally much smaller.
	%In investigation of FFI, there has been numerous searching on electron lepton number (ELN) crossings in core collapse supernova (CCSNe)\cite{PhysRevD.100.043004,Nagakura_2019,PhysRevD.101.023018,PhysRevD.101.043016,PhysRevResearch.2.012046,PhysRevD.101.063001,Abbar_2020,PhysRevD.103.063013,PhysRevD.104.083025,Harada_2022,https://doi.org/10.48550/arxiv.2206.01673}.
	%\mz{Neutrino flavor oscillations, particularly fast flavor conversion, have been studied extensively due to their short timescale and strength compared to the other flavor instability.}
	%\mz{(Please add the following papers with doi: 10.1103/PhysRevD.46.510, 10.1016/0550-3213(93)90175-O, 10.1146/annurev.nucl.012809.104524, 10.1016/j.nuclphysb.2016.02.012, 10.1146/annurev-nucl-102920-050505, 10.48550/arXiv.2207.03561. Refs 1-2 are the earliest papears on the neutrino self-interactions and Refs 3-6 are review papers on collective neutrino oscillation.)}
	%\mz{(If you think the behaviors of FFC is important in this paper, you can here add some sentences about them and cite recent studies on FFC.)}
	The ordinary non-forward scatterings of neutrinos were once thought to destroy the coherence among neutrinos, thus, working against the neutrino flavor oscillations. The interplay
	% flavor conversion. Interplay
	%\mz{the neutrino self-interactions. The interplay}
	between FFC and the ordinary collisions has been investigated from various directions\,\cite{PhysRevD.103.063001,PhysRevD.105.043005,PhysRevD.106.043031,PhysRevD.103.063002,Kato_2021,10.1093/ptep/ptac082,PhysRevD.105.123003,Kato_2022,PhysRevD.106.103031}.
	%\mz{\cite{PhysRevD.103.063001,PhysRevD.105.043005,PhysRevD.106.043031,PhysRevD.103.063002,Kato_2021,10.1093/ptep/ptac082,PhysRevD.105.123003,Kato_2022}.}
	%\mz{(Here I cite 3 papers pointing out scattering-damped FFC and 5 papers showing scattering-enhanced FFC.)}
	%\mz{Contrary to the damping effects, ...}
	Interestingly, some numerical simulations, in which the collision rate was artificially modified by orders of magnitudes, found that FFC can be enhanced by the collisions\,\cite{PhysRevD.103.063002}. It was also pointed out that the collisions may modify the neutrino spectra so that FFC could be driven\,\cite{PhysRevLett.122.091101}. The so-called collisional dilemma, i.e., enhancement or damping of FFC by the ordinary collision, has not been fully resolved so far.\\
	
	%\mz{(Divide into two paragraph before and after. From here, you start to mention about CFI, which is a main theme in this paper. Until here, recent studies on the collisional dilemma between enhancement or damping on FFC... $\to$ Recently, the existence of CFI... I guess it is )}
	%There are regions in core collapse supernova where fast flavor conversion is absent, and in those regions people have been looking for other instability schemes.
	Recently the existence of a new type of flavor conversion referred to as the collisional flavor instability (CFI), which is driven by the collisions themselves and can occur without the NFLN crossing and hence FFC, was pointed out\,\cite{luke2019}. \liu{Properties of CFI were investigated in the same frame work in another paper\,\cite{PhysRevD.106.103031}.} It was also shown that the onset of CFI could be hastened by FFC\,\cite{PhysRevD.106.103029}. 
	%\mz{(You need to introduce CFI in advance through Ref.\,\cite{luke2019} before explaining Ref.\,\cite{PhysRevD.106.103029}.)}
	For homogeneous, isotropic, and monochromatic neutrino distributions the condition for CFI is thought to be that the collision rate for neutrino should be different from that for antineutrino\,\cite{luke2019}. %Collisional flavor instability (CFI) is such an instability that can exist without an angular cross in Electron Leptron Number (ELN), which is the condition for existence of fast flavor conversions, meaning that its occurrence in regions of a core collapse supernovae in absence of fast instabilities is not prohibited.
	%\mz{(It is better that the order of the two sentences above should be reversed. The 2nd sentence gives the conditions of CFI and means that the criteria is outside the conditions of FFC. So, flavor conversion can still occur even in the regions in the absence of FFC. I guess this flow is better?)}
	%\mz{(If you mention the appearance/absence of ELN crossing in CCSNe, you may cite several papers on the ELN search in CCSNe in earlier sentences. e.g., 10.1103/PhysRevD.100.043004, 10.3847/1538-4357/ab4cf2, 10.1103/PhysRevD.101.023018, 10.1103/PhysRevD.101.043016, 10.1103/PhysRevResearch.2.012046, 10.1103/PhysRevD.101.063001, 10.1088/1475-7516/2020/05/027, 10.1103/PhysRevD.103.063013, 10.1103/PhysRevD.104.083025, 10.3847/1538-4357/ac38a0, 10.48550/arXiv.2206.01673.)}
	The existence of CFI was later confirmed numerically for homogeneous, isotropic but non-monochromatic neutrino distributions\,\cite{duan}. The authors added to the criterion for CFI a condition that there exists at least one ELN crossing in neutrino $\it{energy}$. \liu{The first global simulation of CFI was performed for static backgrounds taken from different stages in a core-collapse supernova simulation\,\cite{PhysRevD.107.083016}, demonstrating that the flavor oscillation due to CFI could occur faster than advection.} \liu{On the other hand, the enhancement of CFI by an asymmetry in the collision rates between neutrino and anti-neutrino was first noticed in\,\cite{PhysRevD.106.103031}. The resonance-like behavior was pointed out in\,\cite{2022}.} \\
	
	In the following sections, we derive the exact linear growth rate of the CFI mode with $k =0$ in the \liu{resonance-like structure} based on the dispersion relation for the homogeneous, isotropic and monochromatic neutrino distributions in the background and give the criterion for the \liu{resonance-like phenomenon} as well. We consider not only the isotropy-preserving mode but also the isotropy-breaking mode, which has been somehow overlooked in the literature so far, and demonstrate that they also give rise to the \liu{resonance-like phenomenon}. For the isotropy-preserving mode, on the other hand, we solve the QKE numerically in the \liu{resonance-like region} and study the nonlinear evolution of the system \liu{there}. We show that the saturation is reached more rapidly in that case but that the saturation level is hardly affected.\\
	
	We then procced to the non-monochromatic case. We adopt the  Fermi-Dirac distribution for the neutrino energy. Numerically evaluating the dispersion relation, we find that the growth rate \liu{in} and the criterion for the resonance-like phenomenon obtained in the monochromatic case remain good approximations if the monochromatic collision rates are simply replaced by the mean collision rates. Finally, the effect of non-vanishing wave numbers ($k\neq 0$) in the perturbation or of the anisotropy in the background angular distributions in momentum space is studied again based on the dispersion relation. It is found that CFI \liu{gets weaker in both cases}. We also touch  the case, in which the ELN crossing is present initially and the FFC coexists with the \liu{resonance-like structure of CFI}. %\liu{...and propose a possible scheme on whether collision enhances or suppresses FFI in general. This sentence in the previous version is concerning the overlap of FFI and CFI resonance. The resonance peak shifting effect can apparently enhance or suppress FFI. }

	\section{Dispersion Relation}
	The neutrino flavor content in the two flavor approximation is described by the neutrino flavor density matrix
	\begin{equation}
		\rho(x,P)=
		\left(\begin{array}{cc}
			f_{\nu_e} & S \\
			S^* & f_{\nu_x}
		\end{array}\right),
		\label{denmat}
	\end{equation}
	where the star means the complex conjugate; the diagonal elements are neutrino occupation numbers in the individual flavor eigenstates whereas the off-diagonal elements represent the coherence between the two flavors; $x=(x^{\mu}$) is the spacetime position and $P=(E,\boldsymbol{v})$ is the 4-momentum vector of neutrinos, in which neutrinos are assumed to be ultra-relativistic particles traveling at the speed of light $|\boldsymbol{v}|=1$ in natural units\liu{, which we will employ hereafter through the paper}. We use the signature convention of $\eta_{\mu\nu} = \text{diag}(+1, -1, -1, -1)$ for the Minkowski metric. In the flavor isospin convention, \liu{the} negative energy $E<0$ stand for antineutrinos as $\rho(E)=-\bar{\rho}(-E)$. \liu{Note that} quantities \liu{associated with antineutrinos} are indicated by the bar.\\
	
	The evolution of the flavor density matrix is described by the quantum kinetic equation
	\begin{equation}
		\mathrm{i}v\cdot\partial\rho=\lbrack H,\rho\rbrack+\mathrm{i}C,
		\label{fullqke}
	\end{equation}
	where $H$ is the Hamiltonian
	%\begin{equation}
	%    H=H_{\text{vacuum}}+H_{\text{matter}}+H_{\nu\nu}
	%\end{equation}
	% \mz{$H_{\mathrm{vacuum}}$. Such as `vacuum' in equations are not italic. They are not variables.}
	and $C$ is the collision term. The Hamiltonian has the vacuum, matter and neutrino contributions given as
	% \mz{C is italic.}
	\begin{equation}
		\begin{split}
			&H=H_{\text{vac}}+H_{\text{mat}}+H_{\nu},\\
			&H_{\text{vac}}(x,P)=\frac{M^2}{2E},\\
			&H_{\text{mat}}(x,P)=\sqrt{2}G_\text{F} v\cdot \text{diag}(j_e(x),j_x(x)),\\
			&H_{\nu}(x,P)=\sqrt{2}G_\text{F}v\cdot\int\mathrm{d}P'\rho(x,P')v',\\
		\end{split}
	\end{equation}
	% \mz{It may be a minor detail, but subscript of $G_\mathrm{F}$ is also not italic because its `F' is from Fermi.}
	where $M^2$ is the neutrino mass-squared matrix; $j_\alpha(x)$ is the lepton number 4-current of the charged lepton species $\alpha$; the integral over 4-momentum is abbreviated as
	\begin{equation}
		\int dP=\int_{-\infty}^{\infty}\frac{E^2dE}{2\pi^2}\int\frac{d\boldsymbol{v}}{4\pi}.
	\end{equation}
	The collision term $C$ can be written for neutrinos in \liu{the} relaxation approximation as
	\begin{equation}
		C(x,P)=\frac{1}{2}\{\text{diag}(\Gamma_e(x,P),\Gamma_x(x,P)),\rho_\text{eq}-\rho\},
	\end{equation}
	% \mz{Such as `diag' and `eq' are also not variables.}
	where the curly bracket denotes anti-commutator; $\Gamma_{\alpha}(x,P)$ is the collision rate for the charged lepton $\alpha$; $\rho_\text{eq}$ is \liu{the density matrix for} the equilibrium state that is approached through the collision. The collision term for antineutrinos is written in the same manner.\\
	
	The quantum kinetic equation \ref{fullqke} \liu{may} be linearized with respect to $S$ if $|S| \ll f_i$ as
	\begin{widetext}
		\begin{equation}
			v\cdot(\partial-\Lambda_{0e}+\Lambda_{0x})S_{ex}+
			(f_{\nu_e}-f_{\nu_x})\sqrt{2}G_\text{F}\int\mathrm{d}P'v\cdot v'S_{ex}(P')+
			\frac{1}{2E}\sum_{z=e,x}(M^2_{ez}S_{zx}-S_{ez}M^2_{zx})+\mathrm{i}\Gamma_{ex}S_{ex}=0,
			\label{lineom}
		\end{equation}
	\end{widetext}
	% \mz{The last term is $\mathrm{i}\Gamma_{ex}S_{ex}$?}
	% \mz{You can use \& to align the equations, and you should not use $i$ as an index in summation because imaginary number $\mathrm{i}$ already exists.}
	where $\Lambda_{0z}=\sqrt{2}G_\text{F}[j_{z}(x)+\int\mathrm{d}Pf_{\nu_z}(x,P)v]$ and $\Gamma_{ex}(E)=\left[\Gamma_{e}(E)+\Gamma_{x}(E)\right]/2$.
	% \mz{Subscript $k$ is not adequate here because $k$ is defined as wave solutions later.}
	Assuming the plane wave solution as usual as 
	\begin{equation}
		S(x,P)=S(k,P)e^{\mathrm{i}k\cdot x},
	\end{equation}
	where $k=(\omega,\boldsymbol{k})$ is the 4-wavevector and ignoring the vacuum term\liu{,} which is important for the slow instabilities\,\cite{Airen_2018} but plays a minor role to give initial perturbations for the fast instability, we obtain
	\begin{equation}
		\{v\cdot(k-\Lambda_{0e}+\Lambda_{0x})+\mathrm{i}\Gamma_{ex}\}S_{ex}+(f_{\nu_e}-f_{\nu_x})v\cdot a=0.
	\end{equation}
	where the vector $a$ is defined as
	\begin{equation}
		a=\sqrt{2}G_\text{F}\int\mathrm{d}PS_{ex}(P)v.
		\label{defa}
	\end{equation}
	Then $S$ can be solved as 
	\begin{equation}
		S_{ex}(k,P)=-\frac{(f_{\nu_e}-f_{\nu_x})v\cdot a}{v\cdot(k-\Lambda_{0e}+\Lambda_{0x})+\mathrm{i}\Gamma_{ex}}.
		\label{Sfroma}
	\end{equation}
	Substitution of this expression back into the definition of $a$ in Eq.\,\ref{defa} leads to the following homogeneous equation:
	\begin{equation}
		\Pi_{ex}(k)a_{ex}(k)=0,
		\label{pia0}
	\end{equation}
	where the matrix $ \Pi_{ex}$is defined as
	\begin{equation}
		\Pi_{ex}(k)=\eta+\sqrt{2}G_\text{F}\int\frac{\mathrm{d}P(f_{\nu_e}-f_{\nu_x})v\bigotimes v^\intercal}{v\cdot(k-\Lambda_{0e}+\Lambda_{0x})+\mathrm{i}\Gamma_{ex}}.
	\end{equation}
	$\Lambda$'s in the denominator may be absorbed into $k$, shifting the real part of $k$ alone and unaffecting the instability\,\cite{PhysRevLett.118.021101}. Note that the so-called zero mode with $k=0$ does not mean a mode homogeneous in space after this shift. In the isotropic case the neutrino contribution to the shift vanishes.\\
	
	% \mz{Note that a zero mode ($k=0$) in a co-rotating frame is not always identical to homogeneous mode through the phase shift. Under the isotropic case, the neutrino potential term is reduced to be zero.}
	% \liu{footnote{..} puts the sentences in reference section}
	% \mz{Is `can ignore' correct? Zero mode (k=0) in a co-rotating frame is not always identical to homogeneous mode ($K=0$) through the phase shift. Under the isotropic case, the neutrino potential term is reduced to be zero. Please include a note.}
	Nontrivial solutions of $a$ exist if and only if 
	\begin{equation}
		\det\Pi_{ex}(k)=0,
	\end{equation}
	which gives us the dispersion relation $\omega = \omega(\boldsymbol{k})$. The positive imaginary part of $\omega$, $\mathrm{Im}\,\omega>0$,
	implies that the flavor \liu{eigenstate} is unstable and the perturbation in $S$ grows exponentially in time, an indication of CFI. Introducing the spherical coordinates for the neutrino velocity $\boldsymbol{v}$, we can write down the matrix $\Pi_{ex}$ as
	\begin{widetext}
		\begin{equation}
			\label{pi}
			\begin{split}
				\Pi_{ex}(k)=\eta+\sqrt{2}G_\text{F}\int_{-\infty}^{\infty}\frac{E^2\mathrm{d}E}{2\pi^2}\int_{-1}^{1}\frac{\mathrm{d}c_{\theta}}{2}\int_{0}^{2\pi}\frac{\mathrm{d}\phi}{2\pi}\frac{f_{\nu_e}(E,\boldsymbol{v})-f_{\nu_x}(E,\boldsymbol{v})}{\omega-\boldsymbol{v}\cdot\boldsymbol{k}+\mathrm{i}\Gamma_{ex}(E)}
				\\\times\left(\begin{array}{cccc}
					1 & \sqrt{1-c_{\theta}^2}c_{\phi} & \sqrt{1-c_{\theta}^2}s_{\phi} &c_{\theta} \\
					\sqrt{1-c_{\theta}^2}c_{\phi} & \left(1-c_{\theta}^2\right)c_{\phi}^2 & \left(1-c_{\theta}^2\right)s_{\phi}c_{\phi} & \sqrt{1-c_{\theta}^2}c_{\theta}c_{\phi} \\
					\sqrt{1-c_{\theta}^2}s_{\phi} & \left(1-c_{\theta}^2\right)s_{\phi}c_{\phi} & \left(1-c_{\theta}^2\right)s_{\phi}^2 & \sqrt{1-c_{\theta}^2}c_{\theta}s_{\phi} \\
					c_{\theta} & \sqrt{1-c_{\theta}^2}c_{\theta}c_{\phi} & \sqrt{1-c_{\theta}^2}c_{\theta}s_{\phi} & c_{\theta}^2
				\end{array}\right),
			\end{split}
		\end{equation}
	\end{widetext}
	where the following abbreviations are used: $c_\theta = \mathrm{cos}(\theta), s_\theta = \mathrm{sin}(\theta), c_\phi = \mathrm{cos}(\phi), s_\phi = \mathrm{sin}(\phi)$.\\
	
	% \mz{We often use $\int^{1}_{-1} \mathrm{d}c_{\theta}/2$ instead of $\int \mathrm{d}\theta\,s_{\theta}/2$ here.}
	% \mz{cos \& sin should not be italic in equations. $cos\phi$ means $c\times o\times s\times\phi$. Please rewrite them as $\cos$ \& $\sin$ using predefined commands. The same is for $\det$, $\arg$, $\mathrm{i}$, units, etc...}
	The expression is significantly simplified if the background neutrino is isotropic in momentum space. In fact\liu{,} for $k=0$, $\Pi_{ex}$ becomes even simpler, being diagonal. The equation to be solved is reduced in this case to
	\begin{equation}
		I=\sqrt{2}G_\text{F}\int_{-\infty}^{\infty}\frac{E^2\mathrm{d}E}{2\pi^2}\frac{f_{\nu_e}(E)-f_{\nu_x}(E)}{\omega+\mathrm{i}\Gamma_{ex}(E)}=-1,3.
		\label{k0mode}
	\end{equation}
	Note that the solutions for $I = 3$ are  degenerate with the multiplicity of 3. The solutions of Eq.\,\ref{k0mode} are collectively referred to as the homogeneity-preserving modes although the wave vector is shifted. It should be noted that only the solution branch for $I = -1$ has been  studied in previous papers\,\cite{duan,2022}. This is because the authors of these papers assumed tacitly that the perturbation is isotropic. In fact, the spatial components of $a$ vanish in that case so that the three spatial components of Eq.\,\ref{pia0} become trivial. This is equivalent to ignoring the modes for I = 3. If one allows anisotropic perturbations instead, the spatial components of $a$ become nonzero and the solution branch for $I = 3$ is recovered. For this reason, we call the solution branch for $I = -1$ the isotropy-preserving branch and refer to the solution for $I = 3$ as the isotropy-breaking branch.\\ %\liu{I added the comparison of (16) with Duan's criterion here:}
	
	The analysis of the real and imaginary parts of Eq.\,\ref{k0mode} leads to the necessary condition for the occurrence of CFI, i.e, $\mathrm{Im}[\omega]>0$, that the following function, $F(E)$, of $E(>0)$,
	\begin{widetext}
		\liu{
			\begin{equation}
				\label{ecross}
				\begin{split}
					F(E)=&\bigl[f_{\nu_e}(E)-f_{\nu_x}(E)\bigr]\bigl[\mathrm{Im}\,\omega+\Gamma_{ex}(E)\bigr]\bigl[\bigl(\mathrm{Re}\,\omega\bigr)^2+\bigl(\mathrm{Im}\,\omega+\bar{\Gamma}_{ex}(E)\bigr)^2\bigr]\\
					&-\bigl[f_{\bar{\nu}_e}(E)-f_{\bar{\nu}_x}(E)\bigr]\bigl[\mathrm{Im}\,\omega+\bar{\Gamma}_{ex}(E)\bigr]\bigl[\bigl(\mathrm{Re}\,\omega\bigr)^2+\bigl(\mathrm{Im}\,\omega+\Gamma_{ex}(E)\bigr)^2\bigr]
				\end{split}
			\end{equation}
		}
	\end{widetext}
	%\begin{widetext}
	%\begin{equation}
	%\label{ecross}
	%\begin{split}
	%F(E)=&\bigl[f_{\nu_e}(E)-f_{\nu_x}(E)\bigr]
	%\bigl[\mathrm{Im}\,\omega+\Gamma_{ex}(E)\bigr]
	%\bigl[\bigl(\mathrm{Re}\,\omega\bigr)^2+\bigl(\mathrm{%Im}\,\omega+\bar{\Gamma}_{ex}(E)\bigr)^2\bigr]\\
	%&-\bigl[f_{\bar{\nu}_e}(E)-f_{\bar{\nu}_x(E)\bigr]
	%\bigl[\mathrm{Im}\,\omega+\bar{\Gamma}_{ex}(E)\bigr]
	%\bigl[\bigl(\mathrm{Re}\,\omega\bigr)^2+\bigl(\mathrm{Im}\,\omega+\Gamma_{ex}(E)\bigr)^2\bigr]
	%\end{split}
	%\end{equation}
	%\end{widetext}
	should have at least one zero point, or crossing in energy. When the energy-dependent collision rates, $\Gamma_{ex}(E)$ and $\bar{\Gamma}_{ex}(E)$, are identical for neutrinos and antineutrinos, this condition is reduced to the criterion derived \liu{in} Ref.\,\cite{duan}
	%\liu{Is this the correct reasoning?}
	% \mz{$\to$ I mean where this $+3$ branch is derived from. Previous papers have assumed isotropic distributions at the stage of the QKE so that suppressed the appearance of the $+3$ branch. However, it is not correct and they have missed the additional solutions (actually, they are overwhelmed by the $-1$ branch in the maximum growth rate as you show later). I guess it provides the suggestions that we can consider mainly the $-1$ branch.}
	% \mz{Please mention here that the appearance of the $-1$ branch is consistent with the previous works, Duan's and Xiong's papers, but why the $+3$ branch appears unlike them.}
	\section{Monochromatic Neutrinos}
	\subsection{Exact Solutions in the Linear Analysis}
	In this section, we consider monochromatic neutrinos and solve Eq.\,\ref{k0mode} analytically. We assume the following background distributions:
	\begin{equation}
		f_{\nu_e}(E)-f_{\nu_x}(E)=\frac{2\pi^2}{\sqrt{2}G_\text{F}E^2}[\liu{\mathfrak{g}}\delta(E-\epsilon)-\bar{\liu{\mathfrak{g}}}\delta(E+\bar{\epsilon})],
		\label{monoespe}
	\end{equation}
	where $\liu{\mathfrak{g}},\ \bar{\liu{\mathfrak{g}}},\ \epsilon,\ \bar{\epsilon}$ are model parameters. Then Eq.\,\ref{k0mode} becomes 
	\begin{equation}
		\frac{\liu{\mathfrak{g}}}{\omega+\mathrm{i}\Gamma}-\frac{\bar{\liu{\mathfrak{g}}}}{\omega+\mathrm{i}\bar{\Gamma}}=-1,3,
		\label{monoapp}
	\end{equation}
	where $\Gamma=\Gamma_{ex}(E=\epsilon), \bar{\Gamma}=\bar{\Gamma}_{ex}(E=-\bar{\epsilon})$. %\liu{I added this realistic assumption here:}
	Note that in the typical situation in the supernova core of our concern \liu{following inequalities hold by several orders of magnitude:} $\liu{\mathfrak{g}},\ \bar{\liu{\mathfrak{g}}}\gg \Gamma,\ \bar{\Gamma}$, the fact used in the following. Eq.\,\ref{monoapp} can be solved easily to produce
	\begin{equation}
		\omega_{\pm}=-A-\mathrm{i}\gamma\pm\sqrt{A^2-\alpha^2+\mathrm{i}2G\alpha},
		\label{monoew}
	\end{equation}
	for the isotropy-preserving modes and
	\begin{equation}
		\omega_{\pm}=\frac{A}{3}-\mathrm{i}\gamma\pm\sqrt{\left(\frac{A}{3}\right)^2-\alpha^2-\mathrm{i}\frac{2}{3}G\alpha},
		\label{monoeisob}
	\end{equation}
	for the isotropy-breaking modes. In the above equations, the following notations are introduced:
	\begin{equation}
		G=\frac{\liu{\mathfrak{g}}+\bar{\liu{\mathfrak{g}}}}{2},\ A=\frac{\liu{\mathfrak{g}}-\bar{\liu{\mathfrak{g}}}}{2},\ \gamma=\frac{\Gamma+\bar{\Gamma}}{2},\ \alpha=\frac{\Gamma-\bar{\Gamma}}{2}.
		\label{defGAgmal}
	\end{equation}
	% \mz{Second definition is A?}
	%\liu{The following yellow text will be deleted in the next revision. The exact solutions agree with the criterion of asymmetric collision rates between neutrinos and antineutrinos for the existence of collisional instabilities as pointed out previously\cite{luke2019}. In that paper it was pointed out that the sum of collision rates due to absorption and emission and charged current scattering should be distinct for neutrinos and antrineutrinos.}
	%\liu{Question: what should we learn from Eq. (19)? (Now it becomes Eq. (19) and Eq. (20)) Answer: First of all, the analytical solutions exhibit the resonance, and the resonance mechanism is similar for the the two modes. Then the exact criterion for resonance peak can be derived from the above equation analytically. Further the size of resonance region and the peak shifting behavior due to collision rates can be derived.}
	% \mz{What is the criterion? Please write it here or in the introduction. Some of people can not grasp that yours are consistent with Luke's.}
	%There remains ambiguities in the result of the square root, since there exists two square roots of unity and one is always negative of the other. In this work, we adopt the root with positive imaginary part as the principal value. The solutions can be approximated in two different limits of the parameters as 
	%\liu{And how is it related with John’s criterion? Answer:} 
	Note that when the collision rates are identical between neutrino and antineutrino, or $\alpha=0$, the imaginary part of \liu{$\omega_{\pm}$} is $-\gamma < 0$, implying that they are all stable. It \liu{follows} hence that the inequality of the collision rates is needed for CFI, which was the criterion postulated \liu{in} Ref.\,\cite{luke2019}.\\
	
	The complex frequencies given in Eqs.\,\ref{monoew} and \ref{monoeisob} are approximated as
	\begin{equation}
		\omega_{\pm}=\begin{cases}
			-A-\mathrm{i}\gamma\pm\left(|A|+\mathrm{i}\frac{G|\alpha|}{|A|}\right),& \text{if }A^2\gg G|\alpha|,\\
			-A-\mathrm{i}\gamma\pm\sqrt{\mathrm{i}2G\alpha},& \text{if }A^2\ll G|\alpha|,
		\end{cases}
		\label{isopre1}
	\end{equation}
	for the isotropy-preserving branch and
	\begin{equation}
		\omega_{\pm}=\begin{cases}
			\frac{A}{\liu{3}}-\mathrm{i}\gamma\pm\left(\frac{|A|}{3}-\mathrm{i}\frac{G|\alpha|}{|A|}\right),& \text{if }A^2\gg G|\alpha|,\\
			\frac{A}{3}-\mathrm{i}\gamma\pm\sqrt{-\mathrm{i}\frac{2}{3}G\alpha},& \text{if }A^2\ll G|\alpha|,
		\end{cases}
	\end{equation}
	for the isotropy-breaking branch. %\liu{On the other hand, Eqs. (19), (20) are, I suppose, the same as those derived by Johns and used also by Duan (?). Is it true? If so, we have to mention that here. Answer:} 
	The results for the isotropy-preserving modes were derived previously \liu{outside the resonance-like} region in \cite{luke2019,duan} and \liu{inside it in} \cite{2022}, whereas those for the isotropy-breaking modes have not been presented so far because the authors assumed isotropy not only for the background \liu{but} also for perturbations as we mentioned earlier.\\ %The non-resonance approximation, first equality in Eq.\ref{isopre1}, is equivalent to the one derived by Johns\cite{luke2019}.
	
\begin{figure*}
			\centering
			\subfigure[]{\includegraphics[width=0.49\textwidth]{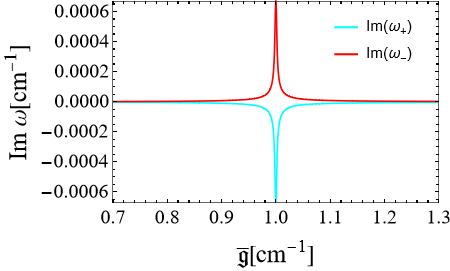}
				\label{monoim}
			}
			\subfigure[]{\includegraphics[width=0.49\textwidth]{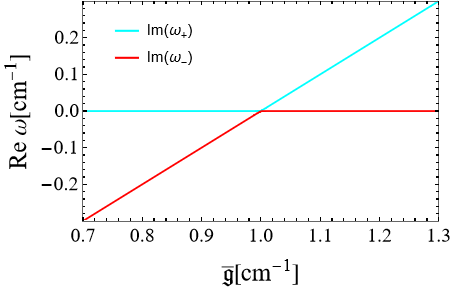}
				\label{monore}
			}
			%     \subfigure[$|C|^{-1/2}(\overline{g})$]{\includegraphics[width=0.49\textwidth]{images/mono3.png}
				%         \label{monoC}
				%     }
			\subfigure[]{\includegraphics[width=0.49\textwidth]{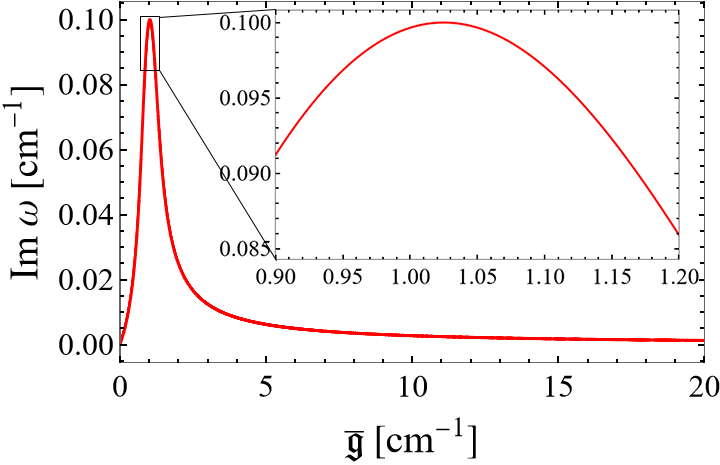}
				\label{monoim2}
			}
			\caption{Real and imaginary parts of the complex frequency $\omega$ for the isotropy-preserving branch as a functio of $\bar{\liu{\mathfrak{g}}}$. For (a) and (b) we take  $\liu{\mathfrak{g}}=1\ \mathrm{cm}^{-1}, \gamma=2.05\times 10^{-7}\ \mathrm{cm}^{-1}, \alpha=-4.5\times 10^{-8}\ \mathrm{cm}^{-1}$, whereas for (c) we artificially magnify the collision rate of antineutrinos to $\bar{\Gamma}=2.5\times 10^{-2}\ \mathrm{cm}^{-1}$, so that $|\alpha|\sim2.5\times 10^{-2}\ \mathrm{cm}^{-1}$ which implies $\bar{\liu{\mathfrak{g}}}\approx 1.025\ \mathrm{cm}^{-1}$ at the \liu{resonance-like} peak. See the text for the notational details.}
			%     Existence of only electron neutrinos and antineutrinos is assumed. The plots (a)-(c) are for the isotropy-preserving branch under the condition $g=1\ \mathrm{cm}^{-1}, \gamma=2.05\times 10^{-7}\ \mathrm{cm}^{-1}, \alpha=-4.5\times 10^{-8}\ \mathrm{cm}^{-1}$, where $\gamma, \alpha$ are the sum and difference of the collision rates of neutrinos and antineutrinos, and $g$ represents neutrino number density converted to unit of $\ \mathrm{cm}^{-1}$ while $\overline{g}$ is left free.
			% \mz{Not $2.05*10^{-7}$, $2.05\times 10^{-7}$.}
			%     As (a) shows, the condition for the resonance part is $|\alpha|\sim|A|=|g-\overline{g}|$ and for the non-resonance part under the parameters of choice is $\overline{g}\not\sim g$. (c) shows the magnitude setting factor multiplied to $\frac{d\omega}{d\overline{g}}$, which shows a roughly flat peak in the resonance part. The other factor, the phase factor, determines whether the slope is positive or negative. Especially the phase factor vanished at $\overline{g}-g=|\alpha|$ corresponding to the enhancement peak. In (d), we artificially magnify the collision rate of antineutrinos to be $\overline{\Gamma}=2.5\times 10^{-2}\ \mathrm{cm}^{-1}$, so that $|\alpha|\sim2.5\times 10^{-2}\ \mathrm{cm}^{-1}$ which implies $\overline{g}_\text{max}\approx 1.025\ \mathrm{cm}^{-1}=g+|\alpha|$. 
			% \mz{Units in Eq.}x
			%     As in (d), the maximum is indeed shifted, the strength is further amplified, and the resonance part is broadened.}
		\label{mono}
	\end{figure*}

\begin{figure*}
		\centering
		\subfigure[]{\includegraphics[width=0.49\textwidth]{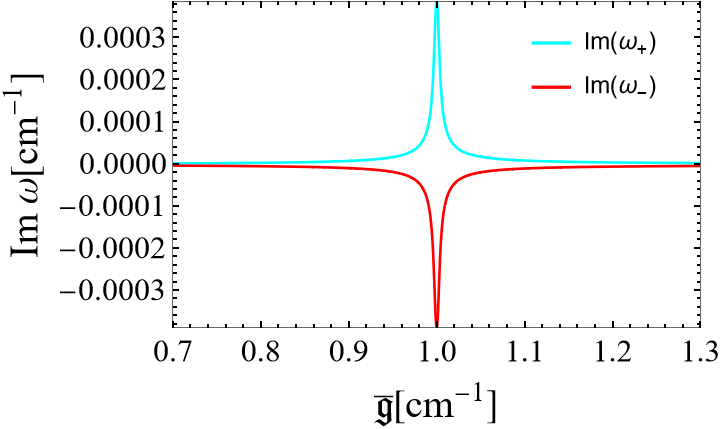}
			\label{monobim}
		}
		\subfigure[]{\includegraphics[width=0.49\textwidth]{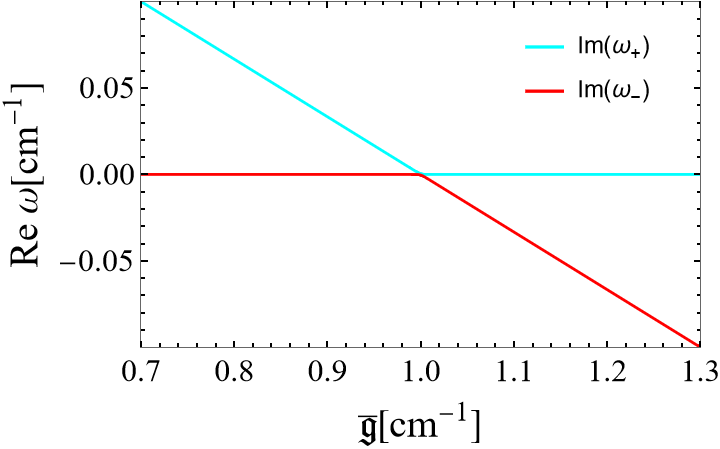}
			\label{monobre}
		}
		\subfigure[]{\includegraphics[width=0.49\textwidth]{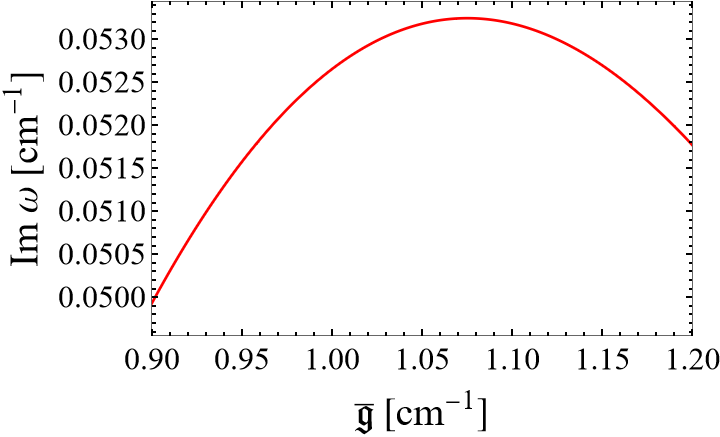}
			\label{monobim2}
		}
		\caption{Same as Fig.\,\ref{mono} but for the isotropy-breaking branch.}
		\label{monob}
	\end{figure*}

	We plot the real and imaginary parts of $\omega$ for the two branches in Figs.\,\ref{mono} and \ref{monob}, respectively, for the following parameter values: $\liu{\mathfrak{g}}=1\,\mathrm{cm}^{-1},\ \gamma=2.05\times 10^{-7}\,\mathrm{cm}^{-1},\ \alpha=-4.5\times 10^{-8}\,\mathrm{cm}^{-1}$. In these plots we take $\bar{\liu{\mathfrak{g}}}$ as a free parameter. It is apparent from the plots of imaginary part that there is a rather narrow region of $\bar{g}$, in which the growth rate is enhanced roughly by two orders of magnitude. This is the \liu{resonance-like structure} that Ref.\,\cite{2022} first pointed out and we will focus on in this paper. In fact, the \liu{resonance-like peak} occurs near the point, at which the real parts of $\omega_{+}$ and $\omega_{-}$ come close to each other. Note that although the \liu{resonance-like structure} was reported only for the isotropy-preserving modes in Ref.\,\cite{2022}, it occurs also for the isotropy-breaking modes as shown in Fig.\,\ref{monob}. 
	\\
	
	To locate where the enhancement occurs exactly, we differentiate $\omega_{+}(\liu{\mathfrak{g}},\ \bar{\liu{\mathfrak{g}}},\ \Gamma,\ \bar{\Gamma})$ of the isotropy-preserving branch in Eq.\,\ref{monoew} with respect to $\bar{\liu{\mathfrak{g}}}$ as
	\begin{equation}
		\begin{split}
			\frac{\partial\omega}{\partial\bar{g}}=&\frac{1}{2}+\frac{1}{2}(A^2-\alpha^2+\mathrm{i}2G\alpha)^{-\frac{1}{2}}(-A+\mathrm{i}\alpha)\\
			=&\frac{1}{2}+\frac{1}{2}|C|^{-\frac{1}{2}}\left[\left(-A\mathrm{cos}\frac{\theta[C]}{2}+\alpha\mathrm{sin}\frac{\theta[C]}{2}\right)\right. +\\ 
			&\left.\mathrm{i}\left(\alpha\mathrm{cos}\frac{\theta[C]}{2}-A\mathrm{sin}\frac{\theta[C]}{2}\right)\right],
		\end{split}
	\end{equation}
	where we define $C$ as 
	\begin{equation}
		C=A^2-\alpha^2+\mathrm{i}2G\alpha=|C|e^{\mathrm{i}\theta[C]}.
	\end{equation}
	%so that
	%\begin{equation}
	%(A^2-\alpha^2+\mathrm{i}2G\alpha)^{-\frac{1}{2}}=|C|^{-1/2}e^{-\mathrm{i}\theta[C]/2}.
	%\end{equation}
	%We define the first quadrant argument 
	%\begin{equation}
	%\theta=\theta\left[|A^2-\alpha^2|+\mathrm{i}|2G\alpha|\right],
	%\end{equation}
	%and the argument in other quadrants can be expressed from it, and the corresponding arguments in second, third, and forth quadrants are $\pi-\theta,\pi+\theta,2\pi-\theta$. Then the derivative becomes 
	%\begin{widetext}
	%\begin{equation}
	% \frac{\partial\omega}{\partial\overline{g}}=\frac{1}{2}+\frac{1}{2}|C|^{-\frac{1}{2}}((-A\mathrm{cos}\frac{\theta[C]}{2}+\alpha\mathrm{sin}\frac{\theta[C]}{2})+\mathrm{i}(\alpha\mathrm{cos}\frac{\theta[C]}{2}-A\mathrm{sin}\frac{\theta[C]}{2}))
	%\frac{\partial\omega}{\partial\overline{g}}=\frac{1}{2}+\frac{1}{2}|C|^{-\frac{1}{2}}\left[\left(-A\mathrm{cos}\frac{\theta[C]}{2}+\alpha\mathrm{sin}\frac{\theta[C]}{2}\right) + \mathrm{i}\left(\alpha\mathrm{cos}\frac{\theta[C]}{2}-A\mathrm{sin}\frac{\theta[C]}{2}\right)\right],
	%\end{equation}
	%\end{widetext}
	% \mz{Please use different bracket size and types. I express the example for the above equation and original one is commented out.}
	The imaginary part of the derivative vanishes at the unique value of $\bar{\liu{\mathfrak{g}}}$ that corresponds to $A^2=\alpha^2$. Thus the \liu{resonance-like peak} occurs when $|A| \approx |\alpha|$, i.e., the number densities of neutrino and antineutrino come close to each other. Note that the difference \liu{between $\Gamma$ and $\bar{\Gamma}$} is usually orders of magnitude smaller than the difference \liu{between $g$ and $\bar{g}$, i.e., }$|\alpha| \ll |A|$.\\%\liu{I changed the description of the inspection of resonance width from the prefacot:} 
	%The resonance width may be inferred from the factor $|C|^{-1/2}$. At the resonance peak, it reaches a plateau of maxima with roughly the same width with the resonance region,
	
	In the isotropy-preserving branch, the imaginary part of $\omega_{+}$ is given approximately as 
	% \mz{Please split them to equation form from the text.}
	\begin{equation}
		\mathrm{Im}\,\omega_{+} \approx\begin{cases}
			-\gamma+\frac{G|\alpha|}{|A|},&\text{if }A^2\gg G|\alpha|,\\ 
			-\gamma+\sqrt{G|\alpha|},&\text{if }A^2\ll G|\alpha|,
		\end{cases}
	\end{equation} 
	%\liu{It is not clear, either, whether the imaginary part is positive indeed. When is it positive? And how is it related with John’s criterion? They should be explained here. Answer:} 
	In the situation of our concern, where $\liu{\mathfrak{g}}, \bar{\liu{\mathfrak{g}}} \gg \Gamma, \bar{\Gamma}$, the first case is non-resonant, since $G\gg|\alpha|$. On the other hand, the second case may fall in the \liu{resonance-like} region. Note that the imaginary part is positive in the second regime unless $\alpha$ is much smaller than $\gamma$. In the first regime, on the other hand, the signature of the imaginary part is determined by the competition of \liu{difference between $g$ and $\bar{g}$ and that between $\Gamma$ and $\bar{\Gamma}$}.\\ %\liu{ust below Eq. (24), you say that the latter is dominant in CCSNe, but (1) what is the latter? The second case in Eq. (24)? Then you should say that the second case is more likely to occur in CCSNe. They cannot be compared with each other, since they are applicable to different regimes. Or did you mean something else? Answer: I meant actually their growth rates. But now this point is transparent so I guess we don't need to say that here explicitly.}
	
	Although we have so far regarded $\omega $ as a function of $\bar{\liu{\mathfrak{g}}}$, one can also take other parameters. If one varies $\liu{\mathfrak{g}}$ instead of $\bar{\liu{\mathfrak{g}}}$, the same criterion $|A| \approx |\alpha| $ is obtained. This \liu{resonance-like} peak, if regarded as a function of one of the collision rates, it reaches the maximum height at a certain value of this rate. Note also that the \liu{resonance-like peak} occurs at $\liu{\mathfrak{g}} \sim \bar{\liu{\mathfrak{g}}}$ but not exactly at $\liu{\mathfrak{g}}=\bar{\liu{\mathfrak{g}}}$. To see this more clearly, we choose a different parameter set, in which we artificially raise the collision rate of antineutrinos to an unrealistically high value $\bar{\Gamma}=0.025\ \mathrm{cm}^{-1}$ so that $|\alpha|\approx 0.025\ \mathrm{cm}^{-1}$. We then obtain  $\bar{\liu{\mathfrak{g}}}\approx 1.025\ \mathrm{cm}^{-1}$ at the peak. This deviation from $\liu{\mathfrak{g}}=1\ \mathrm{cm}^{-1}$ is confirmed in Fig.\,\ref{monoim2}. One should also notice that the \liu{peak} amplitude \liu{in the resonance-like structure} is much larger, and the \liu{peak} width is also much broader in this extreme case. A similar analysis shows that the \liu{resonance-like structure} occurs also for the isotropy-breaking branch. The peak location is a bit different from that for the isotropy-preserving mode and satisfies  $A^2=(3\alpha)^2$, as demonstrated in Fig.\,\ref{monobim2} again for the exaggerated value of $\bar{\Gamma} = 0.025\ \mathrm{cm}^{-1}$. It is also observed that the  \liu{peak} width \liu{in the resonance-like structure} for the isotropy-breaking mode is not appreciably different from that for the isotropy-preserving mode whereas the \liu{peak amplitude} is \liu{smaller} by a factor $\sim2$ for the isotropy-breaking mode than that for the isotropy-preserving mode.\\ %\liu{You have revised up to here last time.}
	
	% \mz{Would it be easier to read if you separate them here with subsections? But I can't think of a good section title.}
	So far we have employed the dispersion relation. The behavior discussed above can be also derived directly from the original quantum kinetic equation. Since it provides a different insight into the \liu{resonance-like structure}, we will look at it below. We assume that the background is homogeneous, isotropic, and monochromatic. The vacuum and matter terms are ignored to focus on CFI. Then the Hamiltonian and the collision term are given as
	\begin{equation}
		\begin{split}
			&H=H_{\nu}=H_\text{d}+H_\text{o}=\\
			&\left(
			\begin{array}{cc}
				\liu{\mathfrak{g}}_{\nu_e}-\liu{\mathfrak{g}}_{\bar{\nu}_e} &  0 \\ 
				0 & \liu{\mathfrak{g}}_{\nu_x}-\liu{\mathfrak{g}}_{\bar{\nu}_x}
			\end{array}\right)
			+
			\left(
			\begin{array}{cc}
				0 &   S-\bar{S} \\
				S^*-\bar{S}^* & 0
			\end{array}\right),
		\end{split}
	\end{equation}
	\begin{equation}
		\begin{split}
			C=\left(
			\begin{array}{cc}
				\Gamma_{e}(\displaystyle \liu{\mathfrak{g}}_{\nu_e,eq}-\liu{\mathfrak{g}}_{\nu_e})& \displaystyle\frac{\Gamma_{e}+\Gamma_{x}}{2}(S_{eq}-S) \\
				\displaystyle\frac{\Gamma_{e}+\Gamma_{x}}{2}(S^*_{eq}-S^*)& \displaystyle\Gamma_{x}(\liu{\mathfrak{g}}_{\nu_x,eq}-\liu{\mathfrak{g}}_{\nu_x})
			\end{array}\right),\\
			\bar{C}=\left(
			\begin{array}{cc}
				\displaystyle\bar{\Gamma}_{e}(\liu{\mathfrak{g}}_{\bar{\nu}_e,eq}-\liu{\mathfrak{g}}_{\bar{\nu}_e})& \displaystyle\frac{\bar{\Gamma}_{e}+\bar{\Gamma}_{x}}{2}(\bar{S}_{eq}-\bar{S}) \\
				\displaystyle\frac{\bar{\Gamma}_{e}+\bar{\Gamma}_{x}}{2}(\bar{S}^*_{eq}-\bar{S}^*)& \displaystyle\bar{\Gamma}_{x}(\liu{\mathfrak{g}}_{\bar{\nu}_x,eq}-\liu{\mathfrak{g}}_{\bar{\nu}_x})
			\end{array}\right).
		\end{split}
	\end{equation}
	Note that the Hamiltonian is divided into the diagonal part $H_\text{d}$ and the off-diagonal part $H_\text{o}$. The equilibrium distributions are set to the unperturbed states:
	\begin{equation}
		\label{equilibrium}
		\begin{split}
			&\liu{\mathfrak{g}}_{\nu_i,\text{eq}}=\liu{\mathfrak{g}}_{\nu_i},\\
			&\liu{\mathfrak{g}}_{\bar{\nu}_i,\text{eq}}=\liu{\mathfrak{g}}_{\bar{\nu}_i},\\
			&\bar{S}^{(*)}_\text{eq}={S}^{(*)}_\text{eq}=0.
		\end{split}
	\end{equation}
	%From the simplified quantum kinetic equation 
	%\begin{equation}
	%\begin{split}
	%&\mathrm{i}\partial_t\rho=[H_{\nu\nu},\rho]+\mathrm{i}C\\
	%&\mathrm{i}\partial_t\overline{\rho}=%[H_{\nu\nu},\overline{\rho}]+\mathrm{i}\overline{C}
	%\end{split}
	%\end{equation}
	%we extract in the linear regime, by assuming occupation numbers being constant, the evolution of flavor coherence.
	The linearized equations for $S$ and $\bar{S}$ are given as
	\begin{equation}
		\mathrm{i}\partial_t\boldsymbol{V}=\boldsymbol{WV},
	\end{equation}
	where $\boldsymbol{V}$ is vector defined as
	\begin{equation}
		\boldsymbol{V}=\left(\begin{array}{cc}
			S\\
			\bar{S} 
		\end{array}\right),
	\end{equation}
	and $\boldsymbol{W}$ is the matrix expressed as
	\begin{equation}
		\left(\begin{array}{cc}
			-(\liu{\mathfrak{g}}_{\bar{\nu}_e}-\liu{\mathfrak{g}}_{\bar{\nu}_x})-\mathrm{i}\frac{1}{2}(\Gamma_e+\Gamma_x)& (\liu{\mathfrak{g}}_{\nu_e}-\liu{\mathfrak{g}}_{\nu_x}) \\
			-(\liu{\mathfrak{g}}_{\bar{\nu}_e}-\liu{\mathfrak{g}}_{\bar{\nu}_x})&(\liu{\mathfrak{g}}_{\nu_e}-\liu{\mathfrak{g}}_{\nu_x})-\mathrm{i}\frac{1}{2}(\bar{\Gamma}_e+\bar{\Gamma}_x)
		\end{array}
		\right).
	\end{equation}
	%\begin{widetext}
	%\begin{equation}
	%    \mathrm{i}\partial_t\left(\begin{array}{cc}
		%         S\\
		%         \overline{S} 
		%    \end{array}\right)
	%    =\\\left(\begin{array}{cc}
		%        -(f_{\overline{\nu}_e}-f_{\overline{\nu}_x})-\mathrm{i}\frac{1}{2}%(\Gamma_e+\Gamma_x)& (f_{\nu_e}-f_{\nu_x}) \\
		%        -(f_{\overline{\nu}_e}-f_{\overline{\nu}_x})&(f_{\nu_e}-f_{\nu_x})-%\mathrm{i}\frac{1}{2}(\overline{\Gamma}_e+\overline{\Gamma}_x)
		%    \end{array}
	%    \right)\left(\begin{array}{cc}
		%         S \\
		%         \overline{S} 
		%    \end{array}\right)\\
	%    \equiv%\mz{\equiv} 
	%    \Lambda V.
	%\end{equation}
	%\end{widetext}
	Note that we assume that the perturbation is also isotropic in deriving these equations here.\\ 
	
	The general solution is given as the superposition of two independent solutions:
	\begin{equation}
		\boldsymbol{V}(t)=\boldsymbol{V}_+e^{-\mathrm{i}\omega_+t}+\boldsymbol{V}_-e^{-\mathrm{i}\omega_-t},
	\end{equation}
	%where the vector 
	%\begin{equation}
	%\boldsymbol{V}(t)=\left(\begin{array}{cc}
		%     S(t)  \\
		%     \overline{S}(t) 
		%\end{array}\right),
		%\end{equation}
		%is defined, and 
		where $\boldsymbol{V}_\pm,\ \omega_\pm$ are the eigenvectors and eigenvalues to the matrix $W$. The eigenvalues are obtained as \cite{duan}
		\begin{equation}
			\omega_\pm%&=-\frac{1}{2}\Bigl[\bigl(f_{\nu_e}-f_{\nu_x}\bigr)-\bigl(f_{\overline{\nu}_e}-f_{\overline{\nu}_x}\bigr)\Bigr]-\frac{\mathrm{i}}{4}\Bigl(\Gamma_e+\Gamma_x+\overline{\Gamma}_e+\overline{\Gamma}_x\Bigr)\pm\Bigl[\bigl[2\bigl((f_{\nu_e}-f_{\nu_x})-(f_{\overline{\nu}_e}-f_{\overline{\nu}_x})\bigr)+\\
			%&\mathrm{i}\bigl(\Gamma_e+\Gamma_x+\overline{\Gamma}_e+\overline{\Gamma}_x\bigr)\bigr]^2-4\bigl[-2\mathrm{i}\bigl(f_{\overline{\nu}_e}-f_{\overline{\nu}_x}\bigr)\bigl(\Gamma_e+\Gamma_x\bigr)+2\mathrm{i}\bigl(f_{\nu_e}-f_{\nu_x}\bigr)\bigl(\overline{\Gamma}_e+\overline{\Gamma}_x\bigr)-\bigl(\Gamma_e+\Gamma_x\bigr)\bigl(\overline{\Gamma}_e+\overline{\Gamma}_x\bigr)\bigr]\Bigr]^{1/2}\\
			=-A-\mathrm{i}\gamma\pm\sqrt{A^2-\alpha^2+2\mathrm{i}G\alpha},
		\end{equation}
		% \mz{Brackets.}
		where the notations %correspond to those used in the dispersion relation as:
		%\begin{equation}
		%    \begin{split}
			%        g&=f_{\nu_e}-f_{\nu_x},\ \overline{g}=f_{\overline{\nu}_e}-f_{\overline{\nu}_x},\\
			%    \Gamma&=\frac{1}{2}(\Gamma_e+\Gamma_x),\ \overline{\Gamma}=\frac{1}{2}%(\overline{\Gamma}_e+\overline{\Gamma}_x),%
			%    \end{split}
		%\end{equation}
		% \mz{Please separate this to equation form from the text. Unreadable.}
		are identical to those for Eq.\,\ref{monoew}.\\
		
		It is interesting to point out that the elimination of the diagonal part of the Hamiltonian does not change the two eigenvalues, which suggests that CFI is driven by an interplay of the off-diagonal part of the Hamiltonian and the collision terms. That may be illuminated more clearly by treating the collision term as a perturbation. Les us first consider the case with $H=H_\text{o}$ and no collision term\liu{,} $C=0$. The eigenvalues and eigenvectors in this case are given as
		\begin{equation}
			\begin{split}
				&\omega_1=0,\ \boldsymbol{V}_1=\left(\begin{array}{cc}
					1 \\
					1
				\end{array}\right),\\
				&\omega_2=-\liu{\mathfrak{g}}-\bar{\liu{\mathfrak{g}}},\ \boldsymbol{V}_2=\left(\begin{array}{cc}
					\liu{\mathfrak{g}} \\
					\bar{\liu{\mathfrak{g}}}
				\end{array}\right).
			\end{split}
			\label{unpert}
		\end{equation}
		where and below we introduce the following notations:
		\begin{equation}
			\begin{split}
				\liu{\mathfrak{g}}&=\liu{\mathfrak{g}}_{\nu_e}-\liu{\mathfrak{g}}_{\nu_x},\ \bar{\liu{\mathfrak{g}}}=\liu{\mathfrak{g}}_{\bar{\nu}_e}-\liu{\mathfrak{g}}_{\bar{\nu}_x},\\
				\Gamma&=\frac{1}{2}(\Gamma_e+\Gamma_x),\ \bar{\Gamma}=\frac{1}{2}(\bar{\Gamma}_e+\bar{\Gamma}_x),
			\end{split}
			\label{notation36}
		\end{equation}
		This is a stable flavor evolution with no growth of amplitudes.\\
		
		Now we reinstate the collision terms but as a perturbation. The characteristic equation is then written as
		\begin{equation}
			\bigl[-\liu{\mathfrak{g}}-(\omega_0+\Delta\omega)-\mathrm{i}\Gamma\bigr] \left[\bar{\liu{\mathfrak{g}}}-(\omega_0+\Delta\omega)-\mathrm{i}\bar{\Gamma}\right]+\liu{\mathfrak{g}}\bar{\liu{\mathfrak{g}}}=0,
		\end{equation}
		where the eigenvalue is expressed as the sum of the unperturbed value given in Eq.\,\ref{unpert} and a (small) shift. To the linear order, the shifts are obtained for the two modes as
		\begin{equation}
			\begin{split}
				&\Delta\omega_1\approx\frac{\mathrm{i}\bar{\liu{\mathfrak{g}}}\Gamma-\mathrm{i}\liu{\mathfrak{g}}\bar{\Gamma}}{\liu{\mathfrak{g}}-\bar{\liu{\mathfrak{g}}}},\\
				&\Delta\omega_2\approx\frac{\mathrm{i}\liu{\mathfrak{g}}\Gamma-\mathrm{i}\bar{\liu{\mathfrak{g}}}\bar{\Gamma}}{\bar{\liu{\mathfrak{g}}}-\liu{\mathfrak{g}}}.
			\end{split}
		\end{equation}
		They are divergent at $\liu{\mathfrak{g}}=\bar{\liu{\mathfrak{g}}}$ unless $\Gamma = \bar{\Gamma}$, an indication of the \liu{resonance-like structure} in the current setting, i.e., the collision term is assumed to be small and so is the shift, the latter of which is no longer correct at \liu{$g \approx \bar{g}$, however}.
		% \mz{It would be neater to put these two equations in the texts.}
		Note that the two shifts have opposite signatures, in qualitative agreement with the exact solution.\\
		
		\subsection{Numerical Simulations}
		%\liu{Revision up to here.}
		% \mz{Would it be easier to read if you separate them here with subsections?}
		Now we go beyond the linear analysis. We solve the quantum kinetic equation
		\begin{equation}
			i\partial_{t}\rho=[H_{\nu},\rho]+iC
			\label{eq39}
		\end{equation}
		for the homogeneous, isotropic, and monochromatic neutrino distributions. The vacuum and matter terms are omitted again. The following simulations are meant to study the nonlinear evolution of the isotropy-preserving mode both \liu{outside and in the resonance-like} regions. For this purpose we vary rather arbitrarily $\bar{\liu{\mathfrak{g}}}$, the distribution function of antineutrino. For comparison, we also run linear simulations, in which the Hamiltonian is fixed to the initial value.\\
		
		Since the vacuum term, which would produce perturbations to the flavor \liu{eigenstate} automatically, is neglected here, we give an initial perturbation by hand as 
		follows: 
		\begin{equation}
			\begin{split}
				&\liu{\mathfrak{g}}=\liu{\mathfrak{g}}_{\nu_e}=1\ \mathrm{cm}^{-1},\\
				&\bar{\liu{\mathfrak{g}}}=\liu{\mathfrak{g}}_{\bar{\nu}_e}\text{ is a free variable},\\
				&S=\bar{S}=10^{-8}+0\mathrm{i}\ \mathrm{cm}^{-1}.
			\end{split}
			\label{monoinitial}
		\end{equation}
		For simplicity, we assume that all neutrinos are initially in the electron flavor and the initial perturbation is isotropic so that the isotropy-breaking mode does not appear in this simulation. 
		%For linear simulations, the initial perturbation is chosen as
		%\begin{equation}
		%\begin{split}
		%    &S=1+0\mathrm{i},\\
		%    &\overline{S}=1+0\mathrm{i}.
		%\end{split}
		%\end{equation}
		%For non-linear simulations, initial perturbation is chosen to be 
		As discussed in the previous section, the occurence of CFI is dictated by the four quantities: $\liu{\mathfrak{g}},\ \bar{\liu{\mathfrak{g}}},\ \Gamma$ and $\bar{\Gamma}$ (see Eq.\,\ref{notation36}).
		%\begin{equation}
		%\begin{split}
		%    &f=f_{\nu_e}-f_{\nu_x},\\
		%    &\overline{f}=f_{\overline{\nu}_e}-f_{\overline{\nu}_x},\\
		%    &\Gamma_{e}+\Gamma_{x},\\
		%    &\overline{\Gamma}_{e}+\overline{\Gamma}_{x}.
		%\end{split}
		%\end{equation}
		%Here $f$'s are number densities in unit of inverse length. The unit is converted to number density in physical volume as 
		%\begin{equation}
		%    \frac{1\mathrm{cm}^{-1}}{\sqrt{2}G_\text{F}\hbar^2c^2}=1.557\times 10^{32}\ \mathrm{cm}^{-3}.
		%\end{equation} 
		%\begin{equation}
		%\begin{split}
		%    &f=f_{\nu_e}=1\ \mathrm{cm}^{-1},\\
		%    &\overline{f}=f_{\overline{\nu}_e}\text{ is a free variable}.
		%\end{split}
		%\end{equation}
		Note that in the linear simulations, $\liu{\mathfrak{g}}$ and $\bar{\liu{\mathfrak{g}}}$
		%\begin{equation}
		%\begin{split}
		%    &f(t)=f,\\
		%    &\overline{f}(t)=\overline{f},
		%\end{split}
		%\end{equation}
		are conserved quantities and unchanged in time.\\
		
		The collision rates are chosen as
		\begin{equation}
			\begin{split}
				&\Gamma_e=\Gamma_x=\frac{\Gamma}{2}=1.6\times 10^{-7}\ \mathrm{cm}^{-1},\\   &\bar{\Gamma}_e=\bar{\Gamma}_x=\frac{\bar{\Gamma}}{2}=2.5\times 10^{-7}\ \mathrm{cm}^{-1},
			\end{split}
		\end{equation}
		which correspond to
		\begin{equation}
			\begin{split}
				&\gamma=2.05\times 10^{-7}\ \mathrm{cm}^{-1},\\
				&\alpha=-4.5\times 10^{-8}\ \mathrm{cm}^{-1}.
			\end{split}
		\end{equation}
		Since the $\Gamma$ and $\bar{\Gamma}$ are much smaller than $\liu{\mathfrak{g}}$ and $\bar{\liu{\mathfrak{g}}}$, the \liu{resonance-like }peak occurs essentially at $\liu{\mathfrak{g}} = \bar{\liu{\mathfrak{g}}}$.\\
		
\begin{figure*}
			\centering
			\subfigure[]{\includegraphics[width=0.49\textwidth]{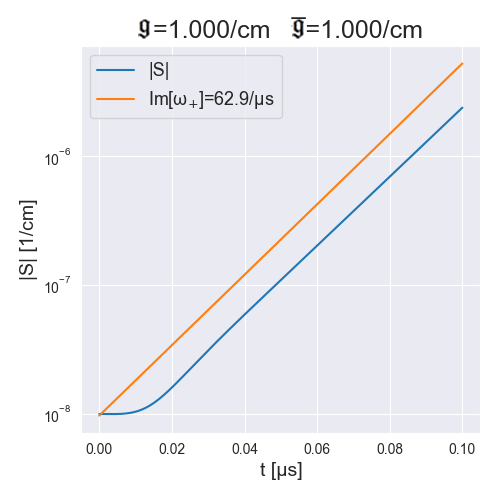}
				\label{monos1}
			}
			\subfigure[]{\includegraphics[width=0.49\textwidth]{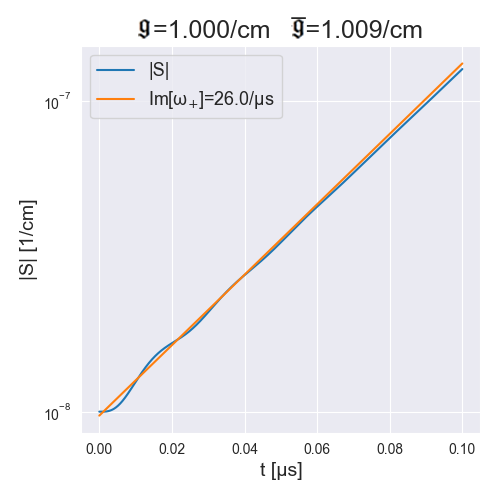}
				\label{monos2}
			}
			\subfigure[]{\includegraphics[width=0.49\textwidth]{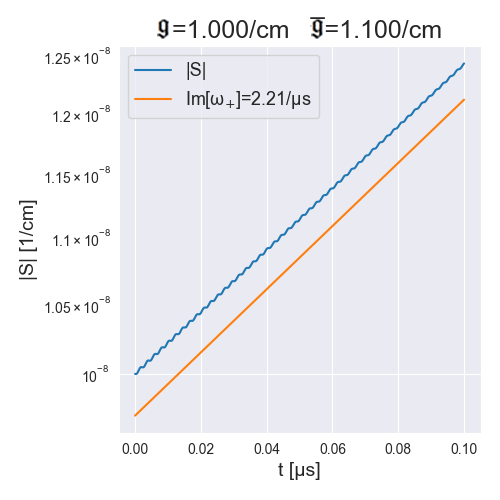}
				\label{monos3}
			}
			\caption{The time evolutions of $|S|$, the modulus of the flavor coherence, for three values of $\bar{\liu{\mathfrak{g}}}$ in the linear simulations. We take $\alpha=-4.5\times 10^{-8}\ \mathrm{cm}^{-1}$ and $\liu{\mathfrak{g}}=1\ \mathrm{cm}^{-1}$. In the plots, the yellow straight lines are the exponential evolution at the growth rate, and the blue curves are numerical \liu{results}. (a) $\bar{\liu{\mathfrak{g}}} = 1.000\ \mathrm{cm}^{-1}$ approoximately corresponds to the \liu{resonance-like} peak, (b) $\bar{\liu{\mathfrak{g}}}=1.009\ \mathrm{cm}^{-1}$ gives a \liu{an} edge \liu{of the resonance-like region,} and (c) $\bar{\liu{\mathfrak{g}}}=1.100\ \mathrm{cm}^{-1}$ is \liu{outside the resonance-like} region. 
				% \mz{For growth rate, please also change $s^{-1}\to\mu s^{-1}$. I think you don't need to display the detailed values. Ex. $\mathrm{Im}(\omega)\approx 2.2\,\mathrm{\mu s^{-1}}$ in the bottom panel. And please space between $t$ and $\mathrm{[\mu s]}$ in the label, that is, $t\,\mathrm{[\mu s]}$.}
			}
			\label{monolin}
		\end{figure*}

		%\begin{equation}
		%\begin{split}
		%    &\Gamma_e=\Gamma_x=\frac{\Gamma}{2}=1.6\times 10^{-7}\ \mathrm{cm}^{-1},\\   &\overline{\Gamma}_e=\overline{\Gamma}_x=\frac{\overline{\Gamma}}{2}=2.5\times 10^{-7}\ \mathrm{cm}^{-1}.
		%\end{split}
		%\end{equation}
		The results of the linear simulations are shown first in Fig.\,\ref{monolin} for three different values of $\bar{\liu{\mathfrak{g}}}$: $1.000\,\mathrm{cm}^{-1}, 1.009\,\mathrm{cm}^{-1}$ and $1.100\ \mathrm{cm}^{-1}$. The first one almost corresponds to the \liu{resonance-like} peak whereas the second and third values give the edge \liu{of} and a \liu{point outside the resonance-like region}, respectively.
		% \liu{I change the order and moved the sentence stating results to the end of this part.}
		% \mz{I guess perhaps readers can not keep up with what you do here. You here perform the time evolution of $S$ `with varying $\bar{f}$', especially around and outside the enhancement. If you demonstrate several situations, you can explain it on the beginning of this paragraph. And then, you provide detailed parameters.}
		The blue curves are the modulus of flavor coherence, $|S|$, plotted against time while the yellow lines indicate for comparison the exponential growths with the values of $\omega_{+}$ given in Eq.\,\ref{monoew} for these settings. As should be clear, $|S|$ shows the exponential growth as expected in all cases. In particular, the growth is much faster indeed in the \liu{resonance-like region}. Note also that the initial conditions are not exactly the \liu{eigenmodes} corresponding to $\omega_{+}$ and some deviations from the exact exponential growth are seen.\\
		
\begin{figure*}
			\centering
			\subfigure[]{\includegraphics[width=0.49\textwidth]{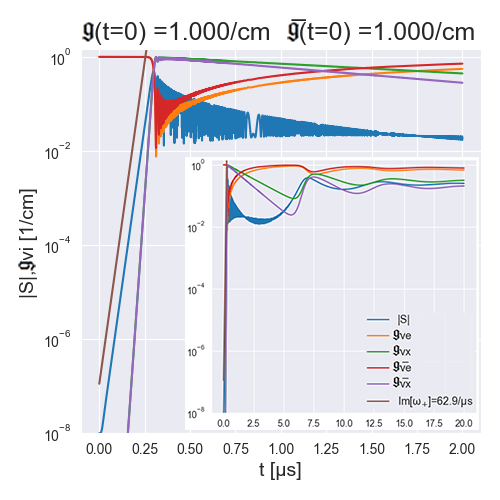}
				\label{monon1}
			}
			\subfigure[]{\includegraphics[width=0.49\textwidth]{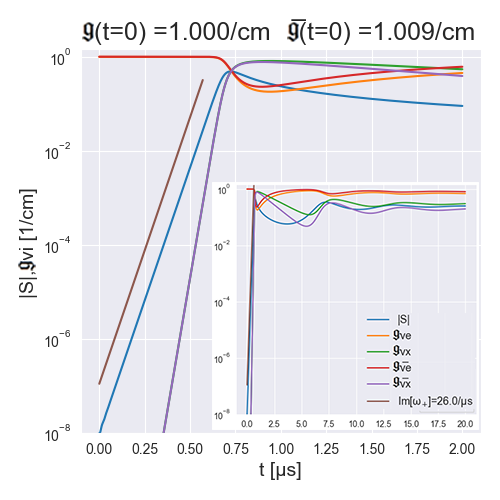}
				\label{monon2}
			}
			\subfigure[]{\includegraphics[width=0.49\textwidth]{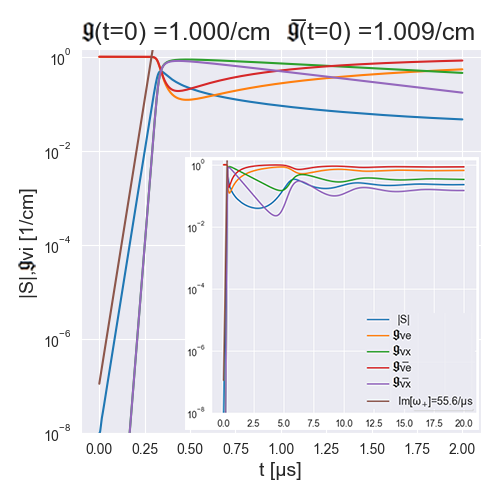}
				\label{monon3}
			}
			%     \subfigure[]{\includegraphics[width=0.49\textwidth]{images/monon4.png}
				%         \label{monon4}
				%     }
			\caption{(a) and (b): the nonlinear time evolutions for the same initial conditions as in Figs.\,\ref{monos1} and \ref{monos2}, respectively. Not only the flavor coherence $|S|$ (blue lines) but also the distribution functions of neutrinos and antineutrinos (colors indicated in the legends) are presented. (c): a variant of (b), in which $\bar{\Gamma}$ is changed to $3.6\times10^{-7}\ \mathrm{cm}^{-1}$ so that the initial condition should be closer to the \liu{resonance-like} peak. \liu{The longer evolutions leading up to the asymptotic states are shown in the insets.}}%\liu{(d): the duration of (a) is extended until an apparent asymptotic state is observed.}}
		\label{monononlin}
	\end{figure*}

		Now we procced to the results of the fully nonlinear simulations run for the first two initial conditions employed in the linear simulations presented just above. % namely
		%\begin{equation}
		%\begin{split}
		%    &\Gamma_e=\Gamma_x=\frac{\Gamma}{2}=1.6*10^{-7}\mathrm{cm}^{-1}\\   &\overline{\Gamma}_e=\overline{\Gamma}=\frac{\overline{\Gamma}}{2}=2.5*10^{-7}\mathrm{cm}^{-1}\\
		%    &f(t=0)=f_{\nu_e}(t=0)=1\mathrm{cm}^{-1}\\
		%    &\overline{f}(t=0)=f_{\overline{\nu}_e}(t=0)\text{ is a free variable}\\
		%    &f_{\nu_x}(t=0)=0\\
		%    &f_{\overline{\nu}_x}(t=0)=0
		%\end{split}
		%\end{equation}
		% \mz{This is the same as the previous conditions, so it is not necessary. Just say you will use the same conditions.}
		They correspond to the peak and edge \liu{of the resonance-like structure}, respectively. In Fig.\,\ref{monononlin} we plot not only $|S|$, the modulus of the flavor coherence and the off-diagonal component of the density matrix, but also the distribution functions of all neutrinos, which are also the diagonal components of the density matrix, as functions of time. For reference, the exponential growths with the values of $\omega_{+}$ in Eq.\,\ref{monoew} for the current settings are again exhibited.\\
		
		One recognizes clearly that there are two distinct phases, the linear phase, in which the flavor coherence grows exponentially at the rate given in the linear analysis, and the nonlinear saturation phase, where the $|S|$ peaks out and levels off thereafter and the distribution functions of neutrinos and antineutrinos are also settled gradually to states that \liu{are steady in the statistical sense and }are different from the initial ones \liu{(see the insets in each panel)}. The transition from the linear phase to the nonlinear phase \liu{may be characterized by $|S| \sim \mathfrak{g}$.} It is found from the result \liu{in} the \liu{resonance-like structure} that the saturation level of the flavor coherence is not particularly large (actually smaller) compared with the case for the edge \liu{of the resonance-like region} and that the role of \liu{resonance-like phenomenon} is just to shorten the time it takes to reach the saturation.\\
		
		This may be supported by another simulation, a variant of the second case, in which the $\bar{\Gamma}$ is changed to $3.6\times10^{-7}\ \mathrm{cm}^{-1}$ with other parameters, particularly $\liu{\mathfrak{g}}$ and $\bar{\liu{\mathfrak{g}}}$, fixed so that it should give the \liu{point closer to the resonance-like} peak. The result is presented as Fig.\,\ref{monon3}. Note that the linear growth rate is twice as large as that in the second case and is close to that in the first case. The saturation occurs earlier consistently with the enhanced linear growth rate. On the other hand, the asymptotic state is almost the same as that in the second case. It is mostly determined by the initial neutrino distributions and little affected by the \liu{resonance-like phenomenon}.\\
		
		\liu{It shold be noted that the asymptotic state may be oscillating in time and is steady not in the literal sense but in the statistical sense. The substantially smaller populations of $\nu_x$ and $\bar{\nu}_x$ in the asymptotic states in these models are the consequence of our choice of the equilibrium state in the relaxation approximation adopted in the simulation and are rather artificial. It should be also mentioned that in reality, where the background matter and neutrinos themselves may change significantly due to advection over the time scale of neutrino oscillations, the asymptotic state may never be reached. It was found that CFI could win the compitition against advection (see \cite{PhysRevD.107.083016}).} What consequences the accelerated saturation may have, if any, for the core-collapse supernova remain to be studied.\\
		%\liu{you proposed the new calculation here:} 
		% \mz{It would be neater to put the equation in the texts.}
		% \mz{Where its 0.1\% comes from? Can we find it from Fig.1(c) for the current parameters? is $\bar{f}=1.009$ still within the enhancement?}
		% \liu{I rewrote these sentences, 1.009 is near the boundary of enhancement.}
		%Since the saturation level of flavor coherence is roughly determined by the initial condition of number densities, we conclude that resonance has no effects on the approximate saturation value of coherence, but only speeds up the rate to saturation, which can be different by at most two orders of magnitudes.
		% \mz{What does the saturation level mean?} 
		
		It is interesting to point out that in the first case corresponding approximately to the \liu{resonance-like} peak, a very fast flavor bouncing occurs in the beginning of the nonlinear saturation phase right after the peak-out. It is due to the diagonal part of the collision term, in which the equilibrium distributions is imposed. In fact, if we reset the equilibrium values to the actual asymptotic values when the nonlinear phase is reached, this feature disappears. %As a matter of fact, in the experimental run with the diagonal part of the Hamiltonian removed artificially we observe the the same evolutions in the linear regime.\liu{I changed this paragraph a little bit.} 
	\section{Non-monochromatic neutrinos}
	We extend the analyses in the previous section to non-monochromatic neutrinos. We assume that neutrinos and antineutrinos have energy spectra given by the Fermi-Dirac distributions. We begin with the linear analysis and discuss the criterion for the \liu{resonance-like phenomenon} as well as the growth rate at the \liu{resonance-like} peak in this case. We then study the nonlinear evolutions of the \liu{system in the resonance-like region} for the isotropy-preserving mode with $k=0$ by numerical simulations. Finally, going back to the linear analysis, we investigate the effects of nonvanishing wave numbers in the perturbation as well as \liu{of} anisotropies in the background on CFI. As an extreme case of the latter, we discuss a possible interplay of the \liu{resonance-like structure} in CFI with FFC on the same footing.\\
	
	\subsection{Linear Analysis}
	We assume that neutrinos and antineutrinos have continuous energy spectra given by the Fermi-Dirac distributions:
	\begin{equation}
		f_i(E,\,g_i,\,T_i,\,\mu_i)=\frac{g_i}{\mathrm{exp}[E/T_i-\mu_i]+1},
		\label{fddist}
	\end{equation}
	where $g_i$, $T_i$ and $\mu_i$ are the model parameters for the neutrino species $i$, and $E$ is the neutrino energy. 
	% \mz{$\mu$ has been already used as $\cos\theta$ in the dispersion relation. Please change either notation. I recommend to use $\mu=c_{\theta}$ and $\cos\phi\to c_{\phi}$ and so on, which simply reduces the volume of the equations, for example. Of course, please formally define them in the case.}
	% \liu{as you suggested things in dispersion relation have been changed and that equation finally fits into one line.}
	Assuming for simplicity that only electron flavor is present initially, we choose
	\begin{equation}
		\begin{split}
			&g_{\nu_e}=1,\\
			&g_{\bar{\nu}_e}\text{ is a free variable},\\
			&g_{\nu_x}=g_{\bar{\nu}_x}=0,
		\end{split}
		\label{eq44}
	\end{equation}
	and the shorthand notation
	\begin{equation}
		\begin{split}
			&g=g_{\nu_e},\\
			&\bar{g}=g_{\bar{\nu}_e}\\
		\end{split}
	\end{equation}
	will be used in the following; the temperatures are set to
	\begin{equation}
		\begin{split}
			&T_{\nu_e}=4\ \mathrm{MeV},\\
			&T_{\bar{\nu}_e}=5\ \mathrm{MeV};
		\end{split}
		\label{Tchoice}
	\end{equation}
	we further assume that $\mu=0$. Although it is not presented here, we confirmed that a non-zero $\mu$ does not change qualitatively the result in the following.
	% \mz{Please space between numbers and units. Ex. $5\,\mathrm{MeV}$.}
	The collision term is now assumed to depend on the energy quadratically as 
	\begin{equation}
		\Gamma(E)=\Gamma_0\left(\frac{E}{10\ \mathrm{MeV}}\right)^2,
		\label{gammae}
	\end{equation}
	with
	\begin{equation}
		\Gamma_0=10^{-5}\ \mathrm{cm}^{-1}
		\label{gm0}
	\end{equation}
	common to all flavors.\\
	
	% \mz{It is better that $\Gamma(E)=\Gamma_0(...)^2$?}
	For later convenience we introduce the number density \liu{multiplied by $\sqrt{2}G_{\mathrm{F}}$}\liu{,} $n_i$, the mean energy\liu{,} $\langle E\rangle_i$, and the mean collision rate\liu{,} $\langle\Gamma\rangle_i$ for the neutrino species $i$ as
	\begin{equation}
		\begin{split}
			n_i&=\liu{\sqrt{2}G_{\mathrm{F}}}\int\frac{E^2\mathrm{d}E}{2\pi^2}f(E,\,g_i,\,T_i,\,\mu_i),\\
			\langle E\rangle_i&=\frac{\liu{\sqrt{2}G_{\mathrm{F}}}}{n_i}\int\frac{E^3\mathrm{d}E}{2\pi^2}f(E,\,g_i,\,T_i,\,\mu_i),\\
			\langle\Gamma\rangle_i&=\frac{\liu{\sqrt{2}G_{\mathrm{F}}}}{n_i}\int\frac{E^2\mathrm{d}E}{2\pi^2}\Gamma(E)f(E,\,g_i,\,T_i,\,\mu_i).
		\end{split}
	\end{equation}
	\liu{These three quantites defined above have same unit.} Note that a change in $g_i$ influences only $n_i$ while a variation in $T_i$ or $\mu_i$ influences all of $n_i,\ \langle E\rangle_i,\ \langle\Gamma\rangle_i$. \liu{We also introduce the following quantities:
		\begin{equation}
			G=\frac{n+\bar{n}}{2},\ A=\frac{n-\bar{n}}{2},\ \gamma=\frac{\langle\Gamma\rangle+\bar{\langle\Gamma\rangle}}{2},\ \alpha=\frac{\langle\Gamma\rangle-\bar{\langle\Gamma\rangle}}{2}.
	\end{equation}}\\
	
	We solve Eq.\,\ref{k0mode} numerically to obtain the dispersion relation. In doing so, the energy range from $0\ \mathrm{MeV}$ to $80\ \mathrm{MeV}$ is divided into nonuniform 128 bins concentrated more at low energies.\\
	
\begin{figure*}
			\subfigure[]{\includegraphics[width=0.49\textwidth]{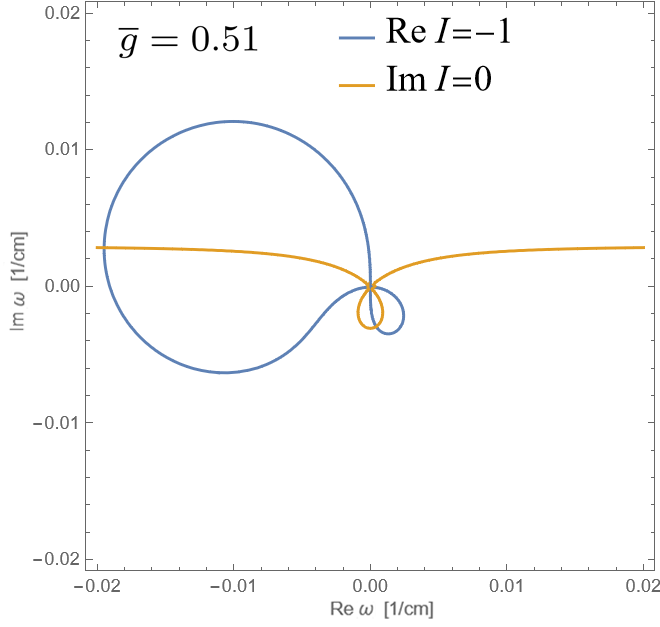}
				\label{dr1}
			}
			\subfigure[]{\includegraphics[width=0.49\textwidth]{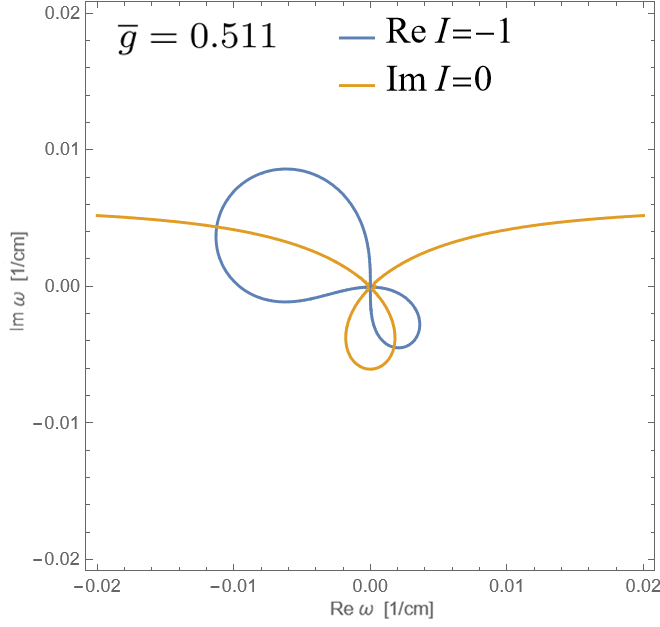}
				\label{dr2}
			}
			\subfigure[]{\includegraphics[width=0.49\textwidth]{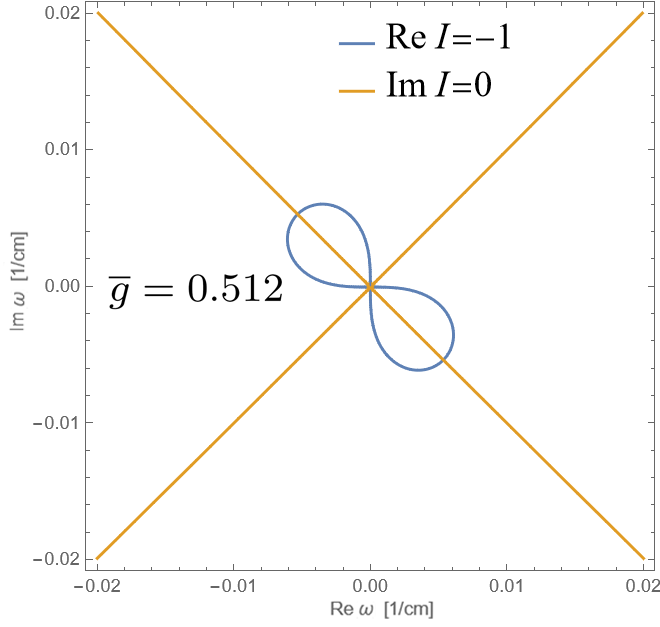}
				\label{dr3}
			}
			\subfigure[]{\includegraphics[width=0.49\textwidth]{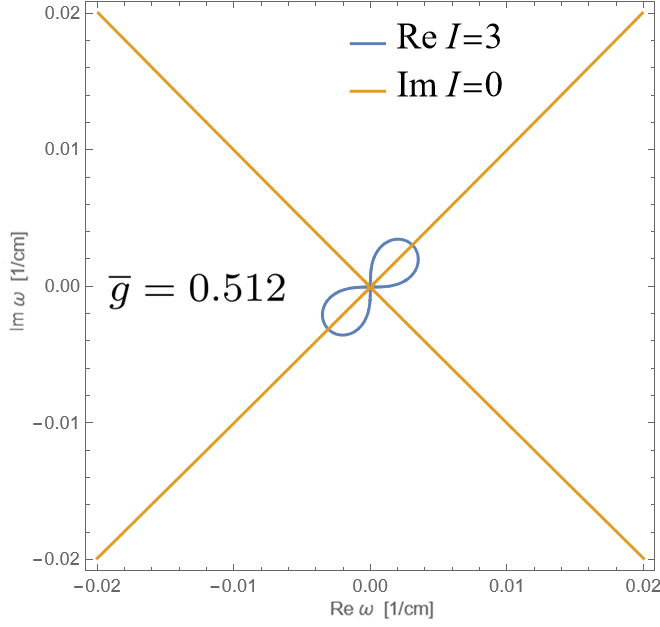}
				\label{dr4}
			}
			\caption{The contours lines for $\mathrm{Re}\ I = -1$ or $3$ (blue) and $\mathrm{Im}\ I = 0$ (orange) in  Eq.\,\ref{k0mode}. Their intersections \liu{give} the dispersion relation  $\omega(k=0)$. Panels (a)-(c) are for the isotropy-preserving branch whereas panel (d) displays the result for the isotropy-breaking branch.}
			\label{cont}
		\end{figure*}

\begin{figure*}
			\centering
			\subfigure[]{\includegraphics[width=0.49\textwidth]{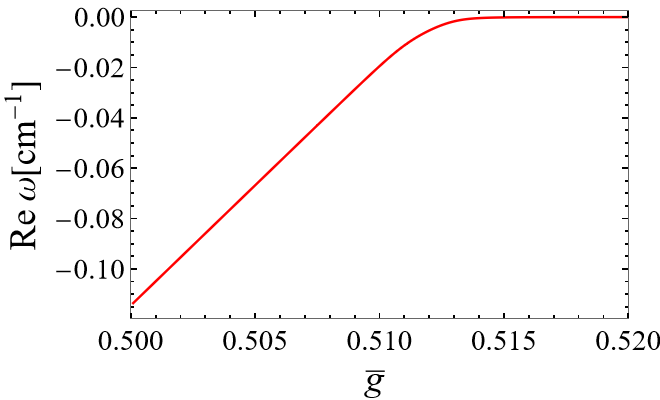}
				\label{wg1}
			}
			%    \subfigure[$\mathrm{Im}\lbrack\omega\rbrack(\overline{g})$]{\includegraphics[width=0.49\textwidth]{images/wg2.png}
				%         \label{wg2}
				%     } 
			\subfigure[]{\includegraphics[width=0.49\textwidth]{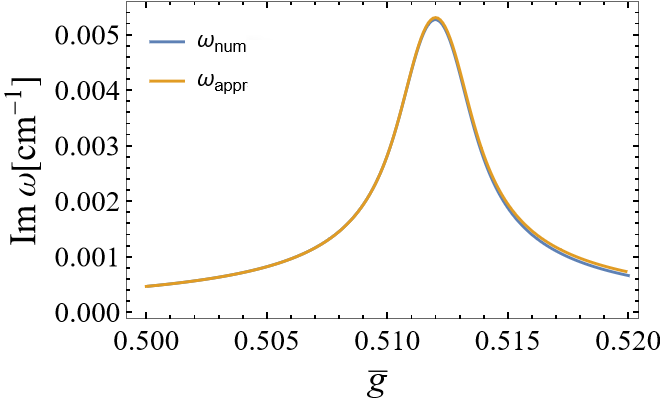}
				\label{wg3}
			}
			\subfigure[]{\includegraphics[width=0.49\textwidth]{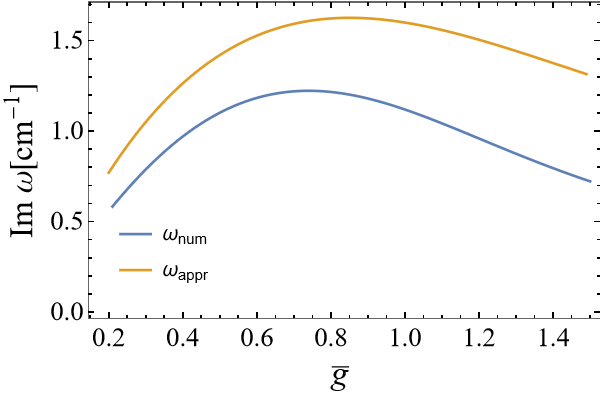}
				\label{wg4}
			}
			\caption{The real (a) and imaginary (b) parts of the complex frequency $\omega$ for the unstable isotropy-preserving  branch as a function of $\bar{g}$. For the imaginary part, the numerical solution $\mathrm{Im}\ \omega_{\text{num}}(\bar{g})$ shown in blue is compared with the approximate one $\mathrm{Im}\ \omega_{\text{appr}}(\bar{g})$ exhibited in orange. In panel (c), the collision rate $\bar{\Gamma}$ of antineutrinos is artificially magnified by $10^5$.}
			\label{wgn}
		\end{figure*}

	We plot in Fig.\,\ref{cont} the contour lines for $\mathrm{Re}\ I = -1$ or $3$ (blue) and $\mathrm{Im}\ I = 0$ (orange) in the complex $\omega$ plane for some representative values of $\bar{g}$. The intersection of the two contour lines gives the dispersion relation $\omega(k)$ at $k=0$. One of the two solutions (except for the origin) with a positive imaginary part is the unstable mode. The first three plots in the figure are for the isotropy-preserving branch at $\bar{g}=0.51,\ 0.511,\ 0.512$, respectively, and the last plot is for the isotropy-breaking branch at $\bar{g}=0.512$.
	% \mz{You can show four figures beyond the columns as $2\times 2$ using }\verb+\begin{figure*}\includegraphics[width=0.5 or less than it \linewidth]{...}\includegraphics[]{} \\ \includegraphics[]{}\includegraphics[]{}\end{figure*}+ \mz{like as `widetext' for equations.}
	For the current choice of parameters, which correspond to
	\begin{equation}
		\begin{split}
			&\langle\Gamma\rangle=2.07031\times 10^{-6}\ \mathrm{cm}^{-1},\\
			&\langle\bar{\Gamma}\rangle=3.23531\times 10^{-6}\ \mathrm{cm}^{-1},\\
			&\alpha=\frac{\langle\Gamma\rangle-\langle\bar{\Gamma}\rangle}{2}\approx-5.825\times 10^{-7}\ \mathrm{cm}^{-1},
		\end{split}    
	\end{equation}
	the \liu{resonance-like} peak is expected to occur approximately at $\bar{g}=0.512$ (panels (c) and (d)) giving $n\approx\bar{n}\approx4.88691\ \mathrm{cm}^{-1}$.
	%$\begin{figure*}
		%    \centering
		%    \label{cont}
		%     \begin{subfigure}[b]{0.3\textwidth}
			%         \centering
			%         \includegraphics[width=0.5\textwidth]{images/dr1.png}
			%         \caption{1}
			%         \label{dr1}
			%     \end{subfigure}
		%     \hfill
		%     \begin{subfigure}[b]{0.3\textwidth}
			%         \centering
			%         \includegraphics[width=0.5\textwidth]{images/dr2.png}
			%         \caption{2}
			%         \label{dr2}
			%     \end{subfigure}
		%     \hfill
		%     \begin{subfigure}[b]{0.3\textwidth}
			%         \centering
			%         \includegraphics[width=0.5\textwidth]{images/dr3.png}
			%         \caption{2}
			%         \label{de3}
			%     \end{subfigure}
		%     \hfill
		%     \begin{subfigure}[b]{0.3\textwidth}
			%         \centering
			%         \includegraphics[width=0.5\textwidth]{images/dr4.png}
			%         \caption{2}
			%         \label{dr4}
			%     \end{subfigure}
		%\end{figure*}
		This is confirmed in Fig.\,\ref{wgn}, where we plot the real and imaginary parts of $\omega$ for the unstable isotropy-preserving branch as a function of $\bar{g}$. One can see a familiar enhancement of the imaginary part, or the growth rate of the instability.\\
		
		We also investigate an extreme case, in which we artificially magnify the collision rate $\bar{\Gamma}$ of antineutrino by five orders of magnitude, $\Gamma_0=1\ \mathrm{cm}^{-1}$ in Eq.\,\ref{gm0}, 
		%\begin{equation}
		%    \bar{\Gamma}(E)=\left(\frac{E}{10\ \mathrm{MeV}}\right)^2\ \mathrm{cm}^{-1},
		%\end{equation}
		while other parameters are unchanged. This gives $\alpha = -1.6174\ \mathrm{cm}^{-1}$. The result is shown in Fig.\,\ref{wg4}. The \liu{resonance-like} peak occurs at $\bar{g}=0.74$, which corresponds to $A=-1.0081\ \mathrm{cm}^{-1}$. The number density of antineutrino is noticably different from that of neutrino in this case just as inferred from the result of the monochromatic case. Again the approximate formula (orange line) is compared with the numerical result shown in blue. This time the approximation is not so successful as before although the peak shift as well as the broadening of \liu{resonance-like region} are captured qualitatively. It is pointed out that thanks to this broadening the case with $n = \bar{n}$ is still in the \liu{resonance-like region}, although not at the peak, and hence the growth rate is enhanced. We explored other neutrino spectra, such as the Fermi-Dirac distributions with non-vanishing chemical potentials or the Gaussian and polynomial distributions that are not too irregular. The results are qualitatively the same as those given above and hence will not be presented here.\\
		
		Finally, it is worth pointing out that the growth rate \liu{of the resonance-like structure} is well approximated by the formula for the monochromatic case, Eq.\,\ref{monoew} for $\omega_{+}$, with the following substitutions:
		\liu{
			\begin{equation}
				\begin{split}
					% &\liu{f} \rightarrow n,\\
					% &\bar{\liu{f}} \rightarrow \bar{n},\\
					% &\Gamma \rightarrow \langle\Gamma\rangle,\\
					% &\bar{\Gamma} \rightarrow \langle\bar{\Gamma}\rangle.
					&\barparent{\liu{\mathfrak{g}}} \rightarrow \barparent{n} \\
					&\barparent{\Gamma} \rightarrow \langle\barparent{\Gamma}\rangle.
				\end{split}
			\end{equation}
		}Note that this approximate formula agrees with the one derived previously in \cite{2022} under the assumption of $|\Gamma|,\ |\bar{\Gamma}| \ll |\omega|$ if $\gamma$ and $\alpha^2$ are neglected accordingly. 
		%\begin{equation}
		%\begin{split}
		%    &G=\frac{n+\overline{n}}{2}\\
		%    &A=\frac{n-\overline{n}}{2}\\
		%    &\gamma=\frac{<\Gamma>+<\overline{\Gamma}>}{2}\\
		%    &\alpha=\frac{<\Gamma>-<\overline{\Gamma}>}{2}
		%\end{split}
		%\end{equation}
		% \mz{I guess these definitions are employed many times in this paper, so you may omit them here.}
		This is demonstrated in Fig.\,\ref{wg3}, where the numerical solution displayed in blue is compared with the approximate one presented in orange. Their agreement matches what was shown in \cite{2022}. Although the criterion for \liu{resonance-like phenomenon} posited in that paper is appropriate only when the collision rates are small, the condition $|A|\sim|\alpha|$ is always satisfied at the \liu{resonance-like peak} regardless of the parameter values.\\
		\subsection{Numerical Simulations}
		We again solve Eq.\,\ref{eq39} numerically for  homogeneous and isotropic neutrinos with the Fermi-Dirac distributions as the energy spectra. The background distributions are actually the same as those employed for the linear analysis in the previous section, Eqs.\,\ref{eq44}-\ref{gm0}. Since the vacuum term is omitted in the kinetic equation, we need to set an initial perturbation by hand, which is also assumed to be isotropic and homogeneous. Hence only the isotropy-preserving branch is considered. We take $\bar{g}$ as a free parameter and vary it so that the unperturbed state \liu{should be} either near the \liu{resonance-like} peak or near the edge \liu{of the resonance-like region}.\\
		
		In the simulation, the energy range from $1\ \mathrm{MeV}$ to $100\ \mathrm{MeV}$ is divided uniformly into 100 bins this time. 
		%\begin{equation}
		%    \Gamma(E)=10^{-5}\left(\frac{E}{10\ \mathrm{MeV}}\right)^2\ %\mathrm{cm}^{-1}.
		%\end{equation}
		%\begin{widetext}
		%\begin{equation}
		%\begin{split}
		%    &f_{\nu_x}(E,t=0)=f_{\bar{\nu}_x}(E,t=0)=0,\\
		%    &f(E)=f_{\nu_e}(E,t=0)=f(E,g=1,T=4\ \mathrm{MeV},\mu=0)=\frac{1}{\mathrm{exp}[E/4\ \mathrm{MeV}]+1},\\
		%    &\bar{f}(E)=f_{\bar{\nu}_e}(E,t=0)=f(E,\bar{g},\bar{T}=5\ \mathrm{MeV},\bar{\mu}=0)=\frac{\bar{g}}{\mathrm{exp}[E/5\ \mathrm{MeV}]+1}
		%\end{split}
		%\end{equation}
		%\end{widetext}
		We adopt the following initial perturbations only to the off-diagonal components of the density matrix
		%\begin{equation}
		%\begin{split}
		%    &S(E)=(1+0.8\mathrm{i})f(E),\\
		%    &\bar{S}(E)=(-0.5+\mathrm{i})\bar{f}(E),
		%\end{split}
		%\end{equation}
		%and for the nonlinear simulation as 
		\begin{equation}
			\begin{split}
				&S(E)=(1+0.8\mathrm{i})10^{-5}f(E),\\
				&\bar{S}(E)=(-0.5+\mathrm{i})10^{-5}\bar{f}(E).
			\end{split}
		\end{equation}
		% \mz{Why are initial perturbations for $S$ and $\bar{S}$ different? Why are they selected?}
		The numbers in the functions are chosen rather arbitrarily.\\
		
\begin{figure*}
			\centering
			\subfigure[]{\includegraphics[width=0.49\textwidth]
				{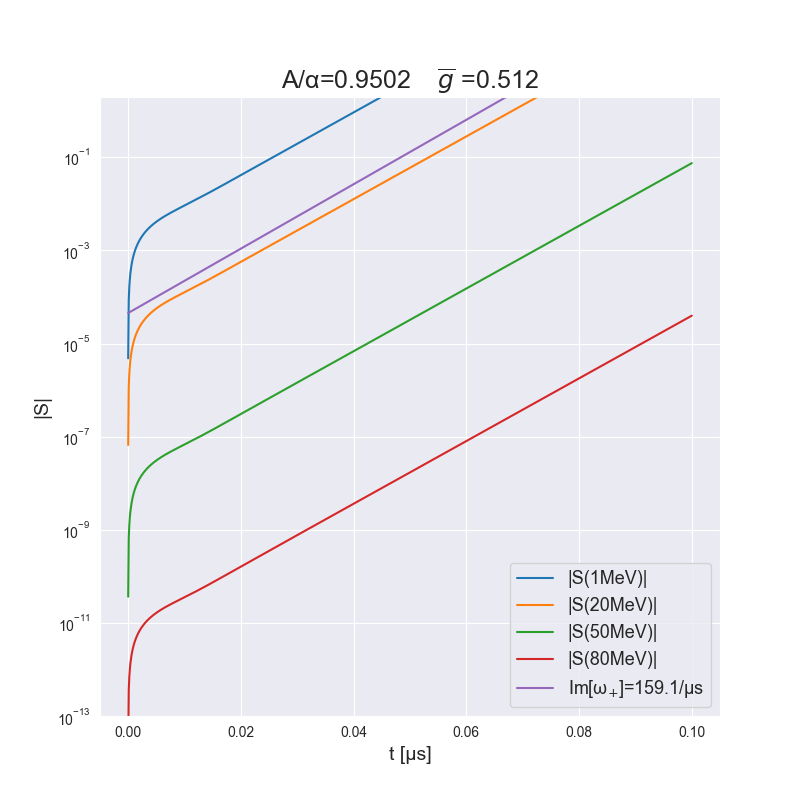}
				\label{fdl1}}
			\subfigure[]{\includegraphics[width=0.49\textwidth]
				{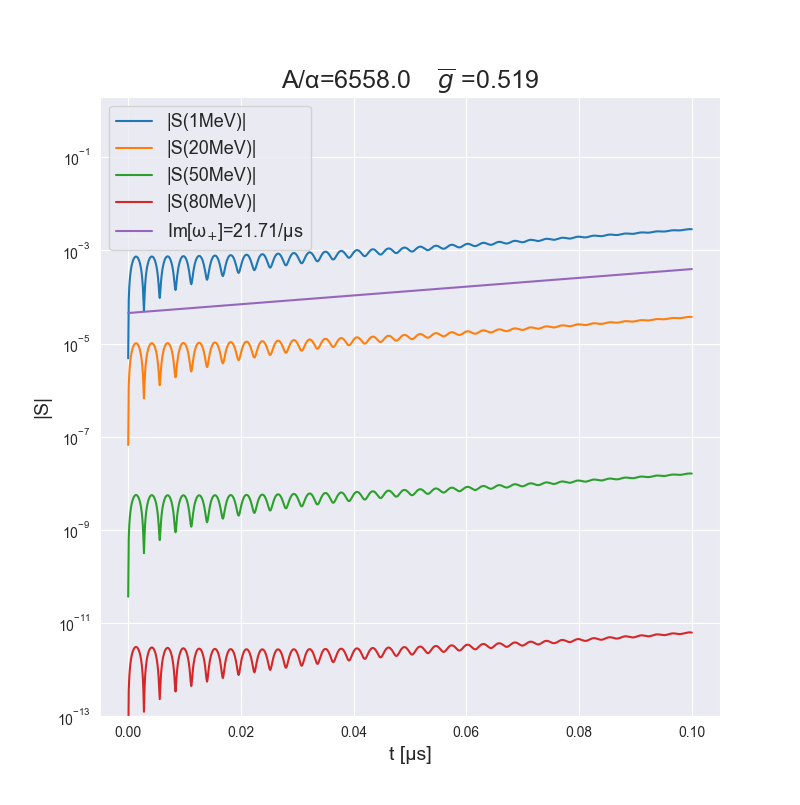}
				\label{fdl2}}
			\caption{The time evolutions of $|S|$, the modulus of the flavor coherence, at different neutrino energies in the linear simulations for the continuous energy spectrum. Two values of $\bar{g}$ are adopted: $\bar{g} = 0.512$ and $0.519$;  $\alpha=-1.165\times 10^{-6}\ \mathrm{cm}^{-1}$, and $g=1$ giving $n_{\nu_e}=4,887\ \mathrm{cm}^{-1}$. The violet lines indicate for reference the exponential evolution at the growth rate given by the linear analysis.}
			\label{fdl}
		\end{figure*}

		The results of the linear simulations, in which the Hamiltonian is fixed to the initial value, is shown in Fig.\,\ref{fdl} for two choices of $\bar{g}$, one corresponding approximately to the \liu{resonance-like} peak and the other giving the edge \liu{of the resonance-like region}. The value of $A/\alpha$ measures how close the initial configuration is to the \liu{resonance-like} peak: the closer to unity it is, the nearer to the \liu{resonance-like} peak the initial state is. 
		%\liu{ref the label of figure doesn't work correctly, it always points to IV, but ref to subfigures works correctly. I am confused.}
		%\mz{$\to$ I corrected the problem. The position of label indicators is wrong. I moved all of them from below the centering to below the caption.}
		%The ratio of difference in number densities to that in mean collision rates $A/\alpha$ roughly indicates whether resonance occurs or not, the closer the value to unity the more enhanced is the system in according to the approximate expression. 
		It is evident that the flavor coherence grows at a common rate for all energies\,\cite{duan}, just as expected from the energy-independent nature of $\omega(k)$. In the \liu{resonance-like region}, it grows much faster indeed.\\
		
\begin{figure*}
			\centering
			\subfigure[]{\includegraphics[width=0.49\textwidth]
				{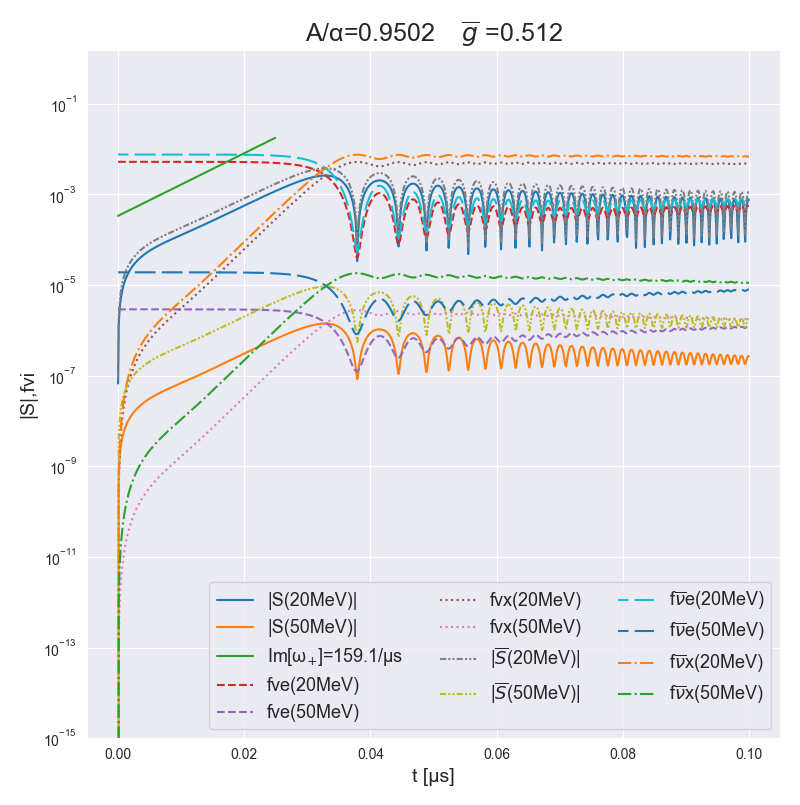}
				\label{fdn1}}
			\subfigure[]{\includegraphics[width=0.49\textwidth]
				{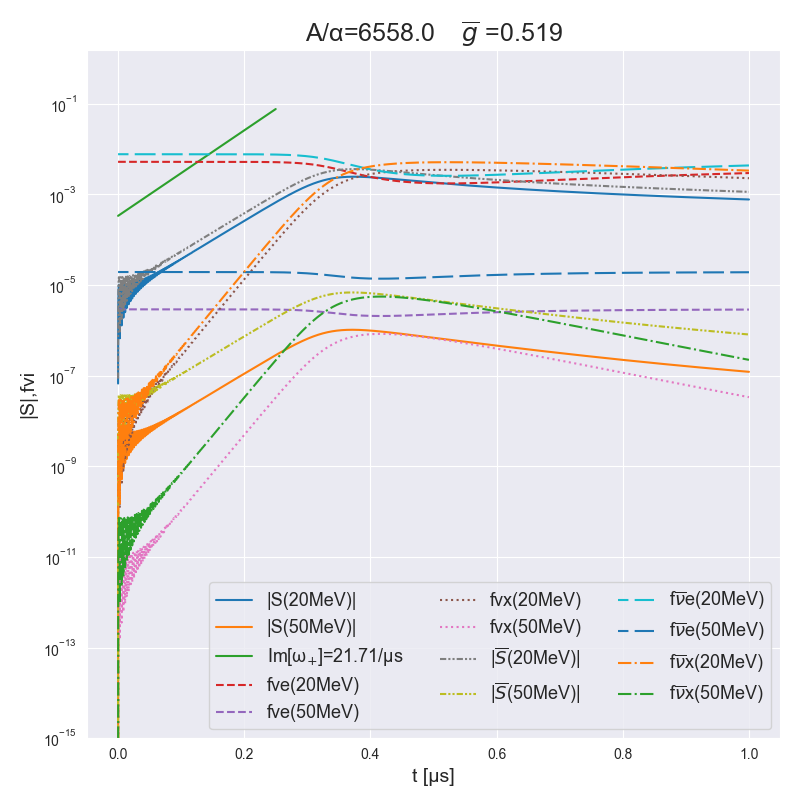}
				\label{fdn2}}
			\caption{The nonlinear time evolutions for the same initial conditions as in Fig.\,\ref{fdl}. Not only the flavor coherence $|S|$ but also the distribution functions of neutrinos and antineutrinos are presented for different energies (\liu{line types} given in the legend). The purple solid line indicates the exponential growth at the rate given by the linear analysis.}
			% \mz{It would be more clearly to read if the legends are arranged in three columns, for example. In the case, the legends for $S,fe,fx$ with the same energy are next to each other.}
			\label{fdn}
		\end{figure*}

\begin{figure*}
			\centering
			\subfigure[]{\includegraphics[width=0.49\textwidth]
				{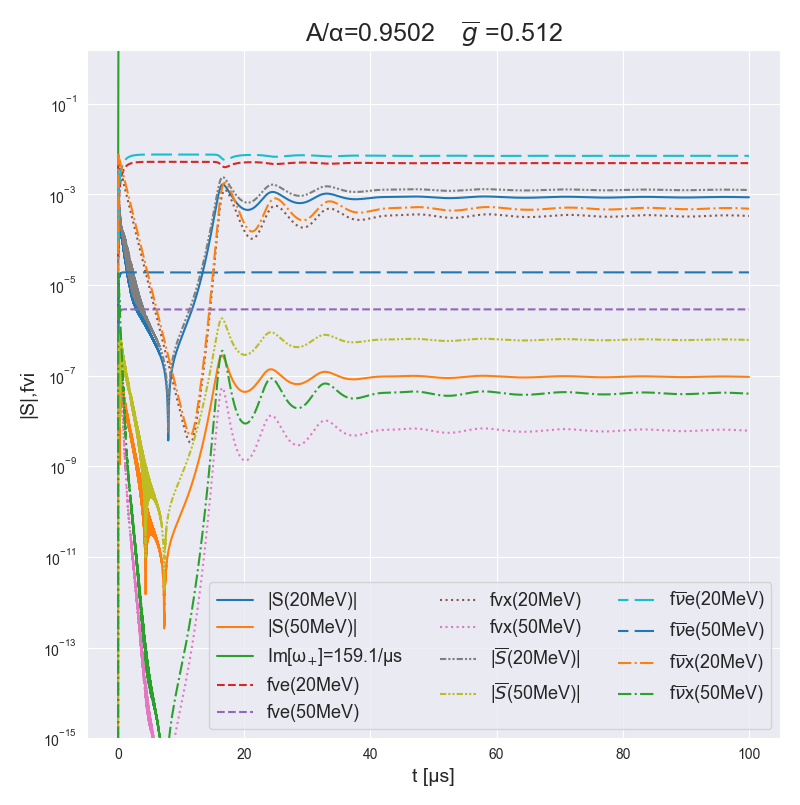}
				\label{lnl1}}
			\subfigure[]{\includegraphics[width=0.49\textwidth]
				{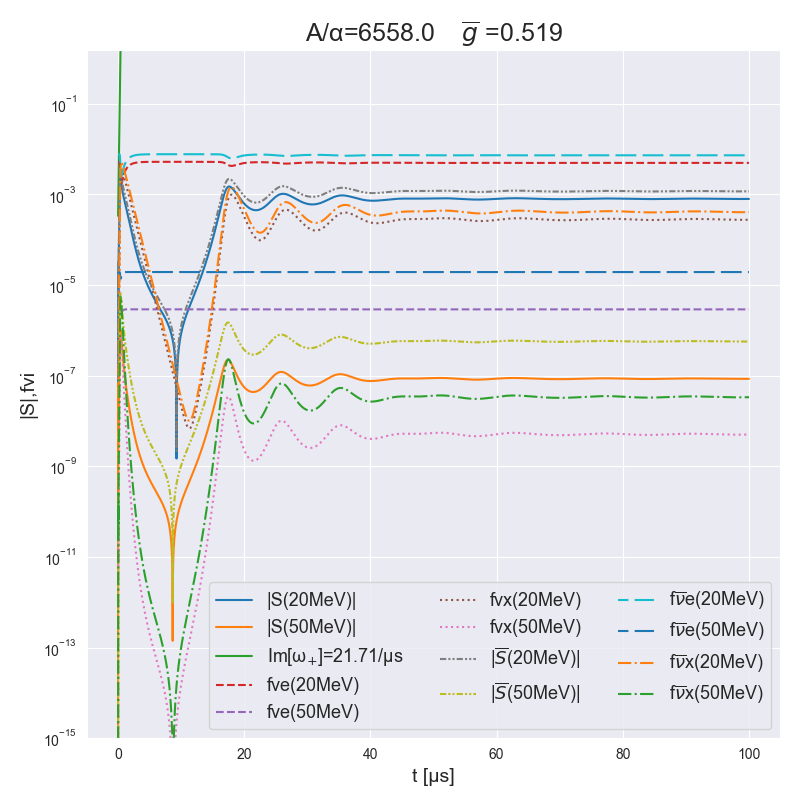}
				\label{lnl2}}
			\caption{\liu{The longer-term evolutions of the same simulations as in Fig.\,\ref{fdn}.}}
			% \mz{It would be more clearly to read if the legends are arranged in three columns, for example. In the case, the legends for $S,fe,fx$ with the same energy are next to each other.}
			\label{lnl}
		\end{figure*}

		The corresponding results for the nonlinear simulations are presented in Figs.\,\ref{fdn} and \ref{lnl}. Again one recognizes that the linear phase is followed by the nonlinear saturation phase also in this case with the non-monochromatic energy spectra. In the linear phase the flavor coherence grows at a common rate for all energies just as in the linear simulations given above \liu{(see Fig.\,\ref{fdn})}. In the case \liu{for the resonance-like peak} (panel (a)), the nonlinear phase is reached much faster than in the \liu{case for the edge of the resonance-like region} (panel (b)).\\
		
		It is also found that the beginning of the nonlinear phase is marked for each energy when $|S(E)| = f(E)$ is satisfied individually, which is a direct extension of the monochromatic case. The saturation level is not much different between \liu{the two} cases \liu{(see Fig.\,\ref{lnl})}, indicating again that the main effect of \liu{the resonance-like phenomenon} is to accelerate the linear growth and the asymptotic state is hardly affected. The bouncing in the nonlinear phase is observed only for the \liu{resonance-like peak}. The bouncing amplitude seems to be related with the difference between the equilibrium distribution and the asymptotic distribution although the exact mechanism of bouncing is not clear at the moment. This is induced by the diagonal part of the collision term. In fact, if we discard the diagonal part effectively by resetting the equilibrium distributions to the asymptotic ones, the bouncing feature is gone just as in the monochromatic case.\\ %\liu{Think about a nonlinear aharmonic oscillator. In the resonance, the system 'relaxes' rapidly toward the convergence, and a leftover decaying oscillation is significant for this case.}
		\subsection{$k\ne 0$ Perturbation/Anisotropy in Background}
		% \mz{Would it be easier to read if you separate them here with subsections?}
		In this last section, we investigate the \liu{resonance-like structure} either for $k \ne 0$ perturbations in the homogeneous and isotropic background or for $k=0$ modes in the homogeneous but anisotropic background. We work on the dispersion relations in the linear regime. The neutrino energy spectra are again assumed to be the Fermi-Dirac distributions given in Eq.\,\ref{fddist}, with the same parameters as in Eqs.\,\ref{eq44}-\ref{Tchoice} and, in addition, with $\bar{g}=g_{\bar{\nu}_e}=0.512$, corresponding to the \liu{resonance-like} peak at $k=0$. The collision rates are given by Eqs.\,\ref{gammae}-\ref{gm0}.\\
		
		We begin with the $k \ne 0$ perturbation in the homogeneous and isotropic background. In the following the wave vector $\boldsymbol{k}$ is assumed to be parallel to the z-axis. We solve Eq.\,\ref{pi} numerically to obtain the dispersion relation. For $k \ne 0$, there appear two nonvanishing off-diagonal components in $\Pi_{ex}$ in addition to the diagonal ones, which survive in the $k\rightarrow0$ limit. In order to see how the isotropy-preserving and isotropy-breaking branches at $k=0$ are mixed with each other to new modes in the $k\ne 0$ case, we decompose $\Pi_{ex}$ as
		\begin{equation}
			\Pi^{\mu\nu}_{ex}=\eta^{\mu\nu}+A^{\mu\nu}.
		\end{equation}
		Then its determinant can be written explicitly in a simple form as
		%\begin{widetext}
		\begin{equation}
			%\begin{split}
			\mathrm{det}\,\Pi_{ex}=(A^{11}-1)^2\left[(A^{00}+1)(A^{33}-1)-(A^{03})^2\right],
			%\end{split}
			\label{newpi}
		\end{equation}
		%\end{widetext}
		where the following relations
		\begin{equation}
			\begin{split}
				&A^{11}=A^{22},\\
				&A^{30}=A^{03},
			\end{split}
		\end{equation}
		are used. Note that in the limit of $k=0$, the following relations hold further:
		\begin{equation}
			\begin{split}
				&A^{30}=A^{03}=0,\\
				&A^{33}=A^{11}=A^{22},
			\end{split}
		\end{equation}
		and the dispersion relation is obtained from 
		\begin{equation}
			\mathrm{det}\,\Pi_{ex} = (A^{00} + 1)(A^{11} - 1)^3 = 0.
		\end{equation} 
		In fact, the isotropy-preserving branch is derived from the first factor and the isotropy-breaking branch is originated from the second factor. In the case of $k \ne 0$, while $A^{11} - 1 = 0$ is unchanged, $(A^{00} + 1)(A^{11} - 1) - (A^{03})^2 = 0$ now mixes the isotropy-preserving and isotropy-breaking branches to produce four branches in general. In the following we look into these modes in detail.\\
		
		From the first factor in Eq.\,\ref{newpi}, a pair of solutions are obtained, which take the following form $a=(0,a_x,a_y,0)$ \liu{and} is perpendicular to $\boldsymbol{k}$, and \liu{they} hence break the isotropy in the $x-y$ plane. On the other hand, the second factor can vanish in three different manners: $(1)A^{00}+1=A^{03}=0,\ (2)A^{33}-1=A^{03}=0$, or $(3)(A^{00}+1)(A^{33}-1)=(A^{03})^2\ne0$. The first case produces a solution of the form: $a=(a_t,0,0,0)$, which is isotropy-preserving. The second case leads to $a=(0,0,0,a_z)$, which is hence isotropy-breaking in the z direction, that is, the direction of $\boldsymbol{k}$. The last one yields a solution with the form of $a=(a_t,0,0,a_z)$ in general, which is also isotropy-breaking. If the first and second factors vanish simultaneously, the solution takes a combined form: for example, the combination of the first factor with case (1) for the second factor \liu{gives} a solution of the following form: $a=(a_t,a_x,a_y,0)$. These cases are exceptional, though, and occur only for special spectra/collision rates. In fact, we find that case (3) is always satisfied in the second factor of Eq.\,\ref{newpi} for the ranges of $k$ and $\bar{g}$ considered in this paper.\\
		
\begin{figure*}
			\centering
			\subfigure[]{\includegraphics[width=0.49\textwidth]{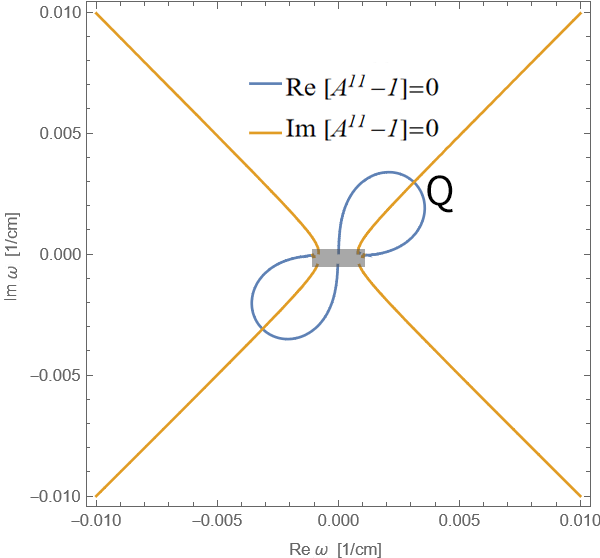}
				\label{fac1dr}}
			\subfigure[]{\includegraphics[width=0.49\textwidth]{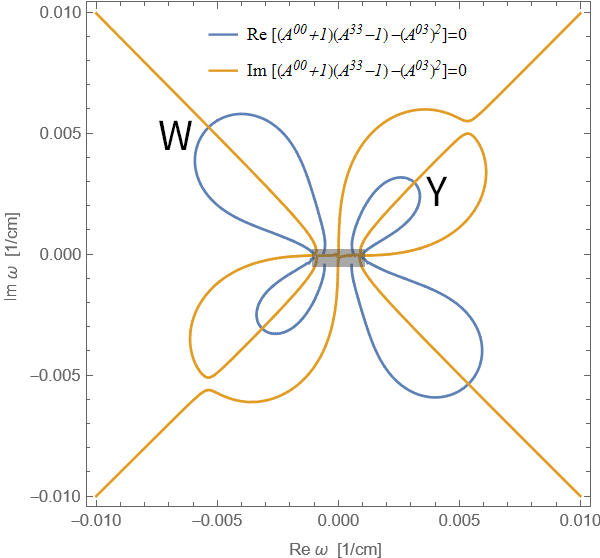}
				\label{fac2dr}}
			\caption{The plots of the contour lines denoted in each panel in the complex $\omega$ plane at $k=0.001\ \mathrm{cm}^{-1}$. \liu{The gray band around the origin is the numerical unreliable region that should be discarded.}}
			\label{facdr}
		\end{figure*}

\begin{figure*}
			\centering
			\subfigure[]{\includegraphics[width=0.49\textwidth]{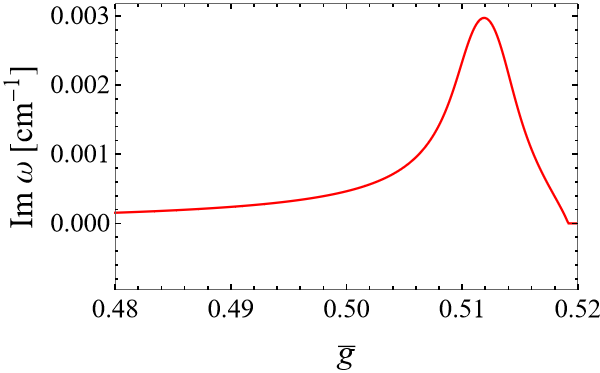}
				\label{kres}}
			\subfigure[]{\includegraphics[width=0.49\textwidth]{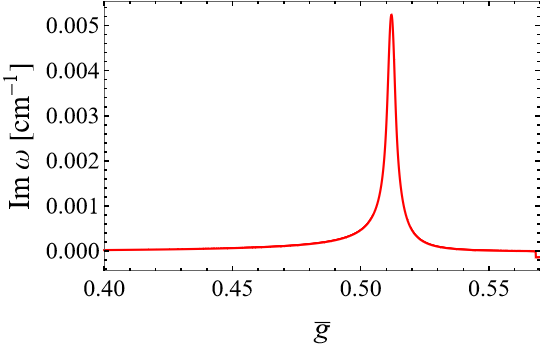}
				\label{krelp}}
			\subfigure[]{\includegraphics[width=0.49\textwidth]{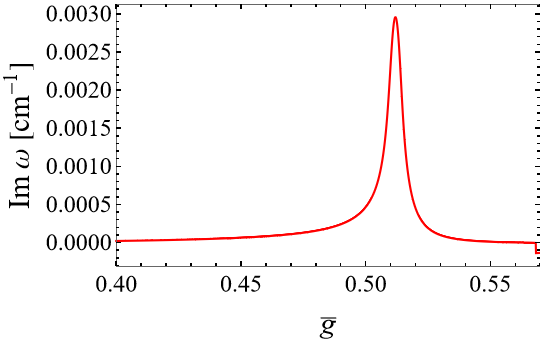}
				\label{krelb}}
			\caption{The \liu{resonance-like structures} for the unstable branches Q in (a), W in (b), and Y in (c) at $k=0.001\,\mathrm{cm}^{-1}$.}
			\label{kreson}
		\end{figure*}

		In Fig.\,\ref{facdr}, we display plots of the contours indicated in each panel for $k=0.001\,\mathrm{cm}^{-1}$. The left panel gives the solutions for the first factor in Eq.\,\ref{newpi} whereas the right panel shows the solutions for the second factor. The latter corresponds to case (3) as mentioned above. At this small $k$, all the modes are not much different from the counterparts at $k = 0$. For later convenience, we refer to the solutions for unstable modes in these plots as Q, W, and Y. Mode Q originated from the first factor in Eq.\,\ref{newpi} merges at $k=0$ with Y from the second factor to give the isotropy-breaking modes. On the other hand, mode W is reduced to the isotropy-preserving mode at $k = 0$. These modes are all isotropy-breaking at $k \ne 0$. In Fig.\,\ref{kreson} we plot the linear growth rates, $\mathrm{Im}\,\omega$, for these three modes as a function of $\bar{g}$ at the same value of $k=0.001\,\mathrm{cm}^{-1}$. The \liu{resonance-like structure} is evident in all cases around $\bar{g} = 0.512$.\\
		
\begin{figure*}
			\centering
			\subfigure[]{\includegraphics[width=0.49\textwidth]{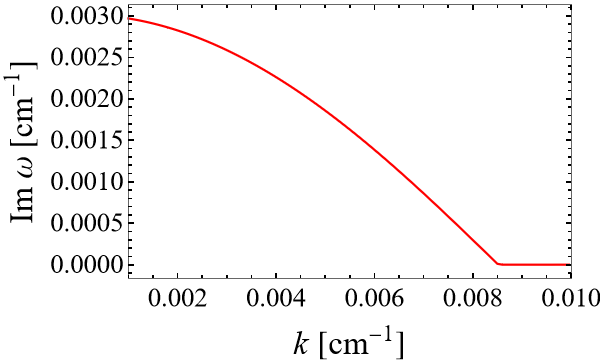}
				\label{facs}}
			\subfigure[]{\includegraphics[width=0.49\textwidth]{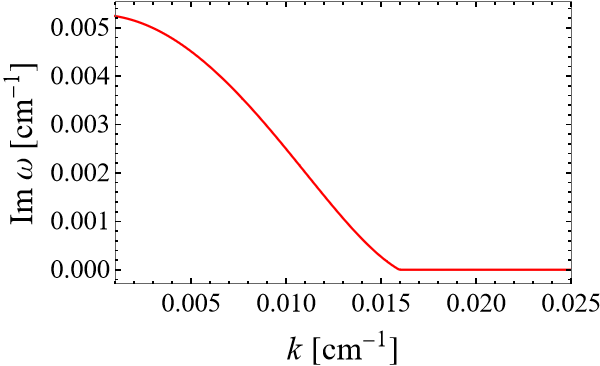}
				\label{faclp}}
			\subfigure[]{\includegraphics[width=0.49\textwidth]{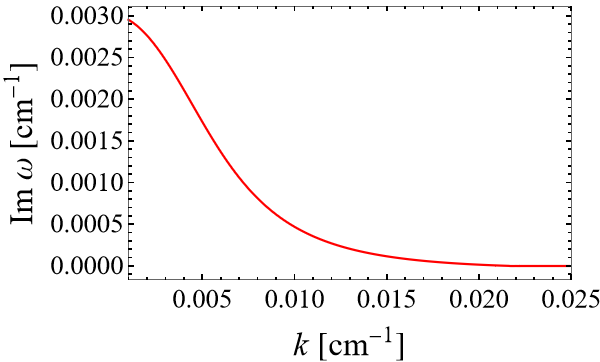}
				\label{faclb}}
			\caption{The linear growth rates for the unstable branches Q in (a), W in (b), and Y in (c) as functions of the wave number $k$.}
			\label{wk}
		\end{figure*}

		In Fig.\,\ref{wk}, we show the $k$-dependence of these growth rates at $\bar{g} = 0.512$, i.e., around the \liu{resonance-like} peak. It is apparent that they decrease monotonically with $k$, indicating that the \liu{resonance-like peak} \liu{gets weaker at} non-vanishing $k$ for all modes. This is actually true \liu{outside the resonance-like} region as well. The non-zero $k$ tends to \liu{reduce} the CFI itself. It is noted that the presence of $\boldsymbol{k}$ in the denominator of Eq.\,\ref{pi} poses a challenge in the numerical integration by discretization in the region $-k<\mathrm{Re}\,\omega<k,\ 0<\mathrm{Im}\,\omega<\Gamma(E_{max})$. We do not think that the numerical solutions in these ranges are reliable and hence do not consider those solutions outside \liu{the resonance-like region or inside it} but \liu{at} large values of $k$ \liu{that approach this problematic region, which is indicated as the gray bands around the origin in Fig.\ref{facdr}.}\\ 
		
		\begin{figure*}
			\centering
			%\subfigure[]{\includegraphics[width=0.49\textwidth]{images/wk.png}
				%\label{wk}}
			\subfigure[]{\includegraphics[width=0.49\textwidth]{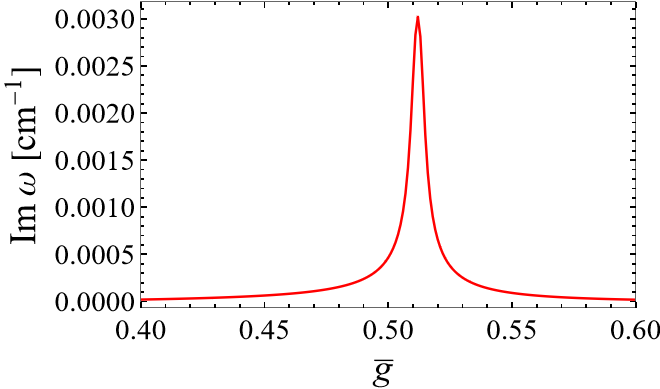}
				\label{cosQ}}
			\subfigure[]{\includegraphics[width=0.49\textwidth]{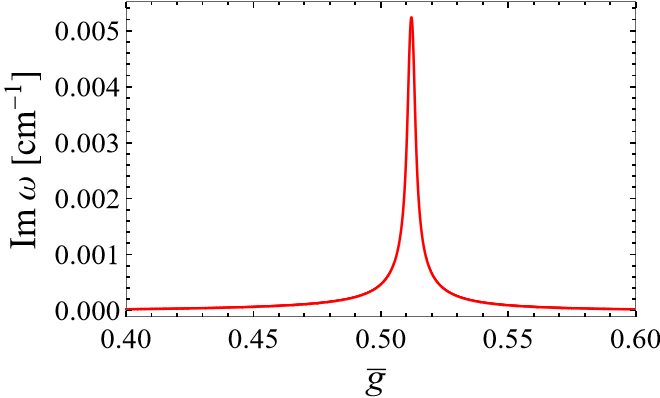}
				\label{cosW}}
			\subfigure[]{\includegraphics[width=0.49\textwidth]{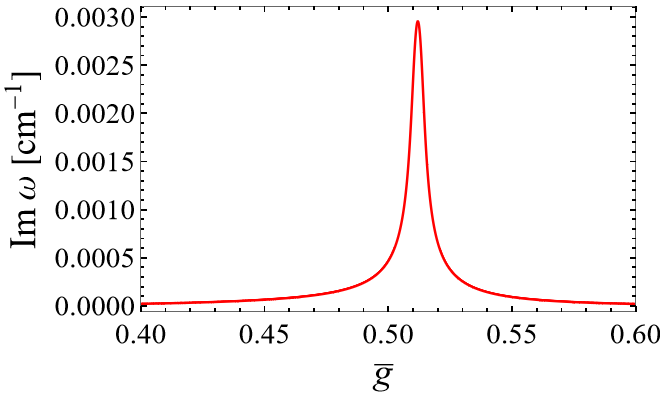}
				\label{cosY}}
			\caption{The \liu{resonance-like structures} for the anisotropic neutrino backgrounds. The degree of anisotropy is identical for $\nu_e$ and $\bar{\nu}_e$: $\delta=\bar{\delta}=0.4$. (a)-(c) correspond to the unstable banches Q, W, and Y respectively. Note that Q and Y are no longer degenerate.}
			\label{cosdep}
		\end{figure*}
		\begin{figure*}
			\centering
			%\subfigure[]{\includegraphics[width=0.49\textwidth]{images/wk.png}
				%\label{wk}}
			\subfigure[]{\includegraphics[width=0.49\textwidth]{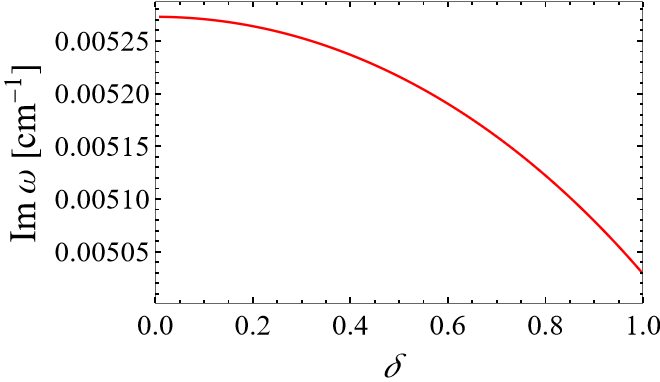}
				\label{wd}}
			\subfigure[]{\includegraphics[width=0.49\textwidth]{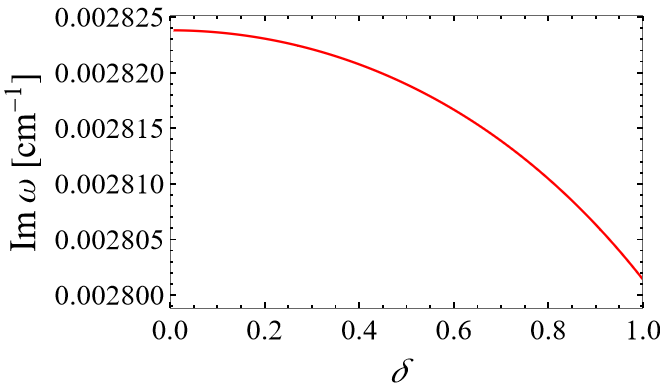}
				\label{wd2}}
			\caption{The linear growth rates for mode W studied as functions of the degree of anisotropy $\delta$. (a) at the \liu{resonance-like} peak ($\bar{g}=0.512$) and (b) at roughly half the peak amplitude ($\bar{g}=0.51$).}
			\label{wdelta}
		\end{figure*}
		Next we study the effect of the anisotropy in the homogeneous neutrino background on the \liu{resonance-like structure}. We introduce the following angular-dependence to the neutrino distributions in momentum space:
		\begin{equation}
			f_i(E,\,g_i,\,T_i,\,\mu_i,\,\theta,\,\delta_i)=(1+\delta_i \mathrm{cos}\theta)f_i(E,\,g_i,\,T_i,\,\mu_i),
		\end{equation}
		where $f_i(E,\,g_i,\,T_i,\,\mu_i)$ on the right hand side is the Fermi-Dirac distribution for neutrino species $i$; $0<\theta<\pi$ is the angle that the neutrino velocity makes with the radially outward direction; the factor $\delta_i$ controls the degree of anisotropy. Since we are interested in the \liu{resonance-like structure} in CFI in this paper, we first consider a case with no angular crossing, and hence no FFC. We then look at two cases with different angular crossings to see the interplay between the \liu{resonance-like structure} in CFI and FFC. We solve Eq.\,\ref{pi} numerically for $k= 0$ to obtain the dispersion relation $\omega (k=0)$ for the anisotropic (but homogeneous) background just described. Note that the integration over the solid angle can be done analytically for $k = 0$.\\
		
		We first present the results for the first case, in which only electron-type neutrinos and antineutrinos are present initially with the same anisotropy: $\delta=\bar{\delta}$. We choose the model parameters as in Eqs.\,\ref{eq44}-\ref{gm0} together with $\bar{g}=0.512$ and $k=0$ so that the unperturbed state corresponds to the \liu{resonance-like} peak at $\delta = 0$. We vary the value of $\delta$ to see how the anisotropy affects CFI in the \liu{resonance-like region}. The choice of $k = 0$ simplifies the analysis a lot. In fact, only $\Pi_{ex}^{03}$, which is linear in $\delta$, is non-vanishing as the off-diagonal components of $\Pi_{ex}$ just as in the previous case with $k \ne 0$, and the diagonal components are unchanged from those for the isotropic case. Since the dispersion relation is obtained from $\mathrm{det}\,\Pi_{ex}\liu{=0}$ given as Eq.\,\ref{newpi} again, we refer to the corresponding modes as Q, W, and Y. Under the current setting, all modes are isotropy-breaking for $\delta\ne0$. Note that mode Q is independent of $\delta$ and $\bar{\delta}$ in this setting, and is always identical to the isotropy-breaking mode for isotropic neutrinos.\\%\liu{I meant that Q is actually $\delta$-independent. Maybe the sentence should be:}\\
		
		In Fig.\,\ref{cosdep} we plot the linear growth rates for the three modes Q, W and Y as functions of $\bar{g}$ at $\delta=0.4$ to show that the \liu{resonance-like structure} occurs indeed also in this case. Note that Q and Y are strictly distinct from each other at $\delta\ne0$. In Fig.\ref{wdelta} we demonstrate how the background anisotropy affects CFI. It is observed that it tends to \liu{reduce} CFI. This time the effect is pretty minor, though.\\
		
		\begin{figure*}
			\centering
			%\subfigure[]{\includegraphics[width=0.49\textwidth]{images/wk.png}
				%\label{wk}}
			\subfigure[]{\includegraphics[width=0.49\textwidth]{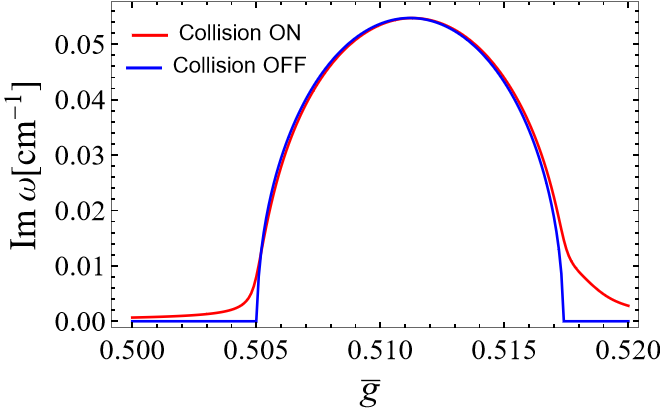}
				\label{woc}}
			\subfigure[]{\includegraphics[width=0.49\textwidth]{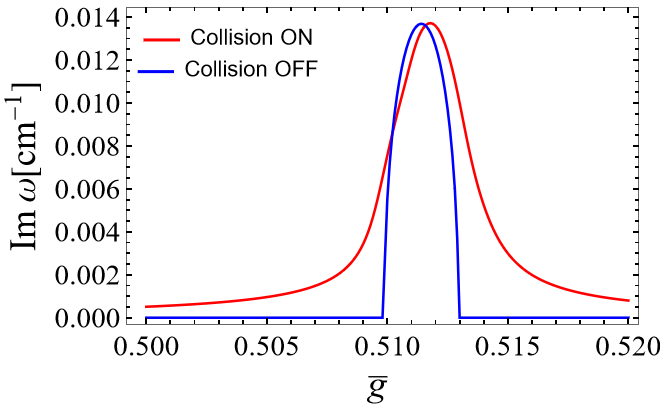}
				\label{interplay}}
			\caption{The linear growth rates as functions of $\bar{g}$ for anisotropic backgrounds with an ELN crossing. The collision term is either turned on (red) or off (blue). (a): $\delta = \liu{0.}1$ and $\bar{\delta} = 0.08$ and (b): $\delta = \liu{0.}1$ and $\bar{\delta} = 0.095$.}
		\end{figure*}
		% \mz{Figure\,\ref{wd} and \ref{woc} are at $k=0$?}\liu{yes they are with homogeneous perturbations}
		Now we go on to the study of a possible interplay of the \liu{resonance-like phenomenon} in CFI with FFC. For this purpose we choose
		\begin{equation}
			\begin{split}
				&\delta=0.1,\\
				&\overline{\delta}=0.08.
			\end{split}
		\end{equation}
		The values of other model parameters are unnchanged from the previous case for $\delta = \bar{\delta}$. We adjust $\bar{g}$ so that there should be an ELN crossing in the angular distributions of $\nu_e$ and $\bar{\nu}_e$. There is indeed an interval of $\bar{g}$, in which the angular ELN crossing and thus FFC occur. In this model, $\nu_e$ is dominant in the radially outward direction while the opposite is true in the inward direction. Note that the mean energies and collision rates for $\bar{\nu}$ are unchanged by the variation of $\bar{g}$. We study at $k=0$ the behavior of the fast-unstable branch in the presence of the collision term.\\
		
		We show in Fig.\,\ref{woc} the growth rates of this mode as functions of $\bar{g}$ both with (red) and without (blue) the collision term. As is evident from the latter, FFC occurs in this mode for $0.505 \lesssim \bar{g} \lesssim 0.517$, where $\mathrm{Im}\,\omega$ is positive. Interestingly, the \liu{resonance-like structure} in CFI takes place almost in the same region as could be inferred from the red line, in which the foot of the \liu{resonance-like} peak can be recognized near the both ends of the FFC range. It is also apparent that the \liu{resonance-like phenomenon} has a very small impact on the growth rate when the FFC is in operation: it only modifies the amplitude very slightly and shifts the peak position only a bit.\\
		
		This is more evident when we take a different parameter set: $\delta=0.1$ and $\bar{\delta}=0.095$. The result is shown in Fig.\,\ref{interplay}. In this case, the degrees of anisotropy are not much different between $\nu_e$ and $\bar{\nu}_e$ and, as a consequence, the range for the ELN crossing is much narrower in $\bar{g}$ (see the blue line). Now the shift in the peak position by the \liu{resonance-like phenomenon in CFI} is apparent. Note that without FFC the height and width \liu{of the resonance-like structure} are essentially unchanged with this variation of $\bar{\delta}$. Because of this shift, the amplitude of FFC, on the other hand, is enhanced or \liu{reduced}, depending on where we look at. Such a shift becomes also noticeable for larger deviations of $\bar{\delta}$ from $\delta$ if the difference in the collision rates is large enough. As mentioned above, it seems that when FFC and the \liu{resonance-like phenomenon} in CFI are simultaneously in operation, the growth rate is set by the former (recall that the growth rate of CFI at the \liu{resonance-like} peak for the current setting at $\delta=\bar{\delta}$ is $\sim0.005\,\mathrm{cm}^{-1}$, i.e., 10 times smaller than the growth rate of FFI alone). More systematic investigations in a broader parameter range are certainly needed \liu{to see how generic this is,} but \liu{they} will be deferred to future studies.
		%\begin{figure}
		%    \centering
		%    \includegraphics[width=0.49\textwidth]{images/interplay.png}
		%    \caption{Similar to Fig.\ref{woc}. The only difference is that the anisotropy is weaker ($\delta=1$ and $\overline{\delta}=0.095$) and so is the fast instability strength. The collisional resonance's effect of peak position shifting is exhibited clearly in this situation, which may lead to seemingly suppression or enhancement of fast instability due to collisional effects in the linear regime.}
		%    \label{interplay}
		%\end{figure}\\
		
		\section{Conclusion}
		We have presented in this paper the results of a systematic study on the \liu{resonance-like structure in} CFI. Employing the two-flavor approximation for simplicity, we have done both linear analysis and nonlinear numerical simulations. The collision is taken into account in the relaxation approximation, which is actually exact for the emission/absorption as well as iso-energetic scatterings of neutrinos. We have always assumed that the neutrino distributions are homogeneous initially but have considered both isotropic and anisotropic distributions in momentum space. We have also taken into account the continuous energy distributions of neutrinos. By changing rather arbitrarily the number densities or the collision rates for the antineutrinos, we have produced both \liu{configurations in and out of the resonance-like region} freely.\\
		
		Starting with the linear analysis of the simplest case, i.e., the monochromatic, homogeneous and isotropic background with $k=0$ perturbations, we have analytically obtained the dispersion relations both for the isotropy-preserving and isotropy-breaking modes. Note that the latter has been overlooked in the literature so far. We have confirmed that the \liu{resonance-like structure} shows up at $A \sim \alpha$ (see Eq.\,\ref{defGAgmal} for \liu{the} notations) for the isotropy-preserving branch, where the two modes, $\omega_{+}$ and $\omega_{-}$ (Eq.\,\ref{monoew}), come close to each other. We have demonstrated that a similar feature occurs at $A \sim 3 \alpha$ also for the isotropy-breaking branch. In both cases, the \liu{resonance-like} peak obtains at $\liu{\mathfrak{g}} \sim \bar{\liu{\mathfrak{g}}}$ (Eq.\,\ref{monoespe}) but not exactly at $\liu{\mathfrak{g}} = \bar{\liu{\mathfrak{g}}}$. In fact, the deviation becomes larger for a greater difference between the collision rates, $\Gamma$ for neutrino and $\bar{\Gamma}$ for antineutrinos. The \liu{resonance-like region} is also broadened in that case.\\
		
		We have then conducted numerical simulations for the same background setting to investigate the nonlinear evolutions of the perturbations, which were again assumed to have $k = 0$ and added only to the off-diagonal components of the density matrix. We have confirmed the exponential growth at the rate given by the linear analysis in the linear phase, which we have found is followed by the nonlinear saturation phase, where the flavor coherence levels off and the distribution functions are settled to new steady states asymptotically. We have observed a bouncing with large amplitudes only \liu{around} the \liu{resonance-like peak}. Its mechanism is unclear for the moment but we have demonstrated that it is induced mainly by the diagonal part of the collision term.\\%\liu{In all simulations the matter background are assumed to be static over the time scale used, which may not be realistic in a typical supernova.}
		
		We have then proceeded to the non-monochromatic case, in which we assumed that neutrinos have Fermi-Dirac distributions as their energy spectra. In the linear analysis for the homogeneous and isotropic background with the $k=0$ perturbation, we have shown that there are again isotropy-preserving and isotropy-breaking branches and that both of them give \liu{resonance-like structure}. We have also demonstrated that the growth rate \liu{in the resonance-like structures} is well approximated by the exact formula for the monochromatic case with an appropriate substitution of variables as long as the collision rates are not much different between neutrino and antineutrino. This was pointed out in the previous work\,\cite{2022} for small collision rates. Our results have extended its validity.\\
		
		The nonlinear simulations have been done for the same background as for the linear analysis above. The perturbation was assumed to have $k = 0$ and to be also isotropic. Only the isotropy-preserving mode has been hence calculated. As expected from the dispersion relation, we have observed that the perturbation grows exponentially at the same rate for all energies of neutrinos. We have found that for different energies of neutrinos the nonlinear saturation phase begins when their flavor coherence becomes of the \liu{similar} amplitude \liu{to} the distribution functions at their energies, $|S(E)| \sim f(E)$ (Eq.\,\ref{denmat}), a direct extension of the monochromatic case. We have found that the saturation level is not much affected by the \liu{resonance-like phenomenon} and its main role is to shorten the time it takes to get to the saturation. We have also seen the bouncing of the flavor coherence after its peak-out \liu{near} the \liu{resonance-like peak} alone again.\\
		
		We finally conducted the linear analysis either for the $k \ne 0$ perturbation to the isotropic background or for the $k=0$ perturbation to anisotropic background configurations. In the former we have shown how the isotropy-preserving and isotropy-breaking branches at $k = 0$ are mixed for $k \ne 0$ to produce new branches and demonstrated that \liu{resonance-like structure} occurs just in the same way for all these modes. We have also found that \liu{the CFI is weaker for modes with $k\ne0$ than $k=0$}.\\ 
		
		For the anisotropic background with no electron-lepton-number, or ELN, crossings, we have shown that \liu{resonance-like structure} shows up again in the same way and that the \liu{CFI} tends to \liu{be weaker, albeit slightly, for the anisotropic background}. \liu{With} an ELN crossing, on the other hand, the growth rate \liu{seems to be} set by the fast flavor conversion, or FFC, and the \liu{resonance-like phenomenon} in CFI, %\liu{! Question: is this generic or specific to a particular setting? ! Answer: This is equivalent to the question that whether FFC maximizes at exactly/or near equal number densities. Consider the angular-dependent difference in energy-integrated number densities D(angle)=f(angle)-fbar(angle). It seems that FFC is stronger if the region D>0 and D<0 are larger and similar. Then that implies equal total number densities, distributed differently to angles, which is also the criterion for CFI resonance in weak collisions.} 
		the region of which tends to overlap with the range of FFC, \liu{shifts} the peak position and \liu{broadens} the range of instability. The growth rate may be enhanced or \liu{reduced}, depending on the position in the unstable range.\\
		
		There remain many issues to be addressed further. Not to mention, we need to extend the analysis to three flavors. The relaxation approximation should be removed to incorporate non-isoenergetic collisions. The parameter regions considered in this paper are rather limited. For instance, we have invetigated rather small $k$ in the $k \ne 0$ perturbation although new modes with different properties may emerge for larger $k$. Interplays of the \liu{resonance-like phenomenon} in CFI with FFC in the nonlinear phase should be studied, probably numerically. Indeed the growth of the isotropy-breaking modes may induce FFC in that phase. The mechanism of the large-amplitude bouncing in the \liu{resonance-like peak} case needs to be understood. The asymptotic state in the nonlinear saturation phase should be characterized. Eventually we are interested in what consequences, if any, the \liu{resonance-like structure} in CFI may have for core-collapse supernovae and compact object mergers. \liu{In fact, the advection, which is ignored in this paper, occurs in these realistic situations and the build-up time of flavor instabilities becomes crucially important\,\cite{PhysRevD.107.083016}. }They \liu{all} will be future works.
		\section{Acknowledgement}
		J.L. thanks Taiki Morinaga for introducing this field of research to him and providing well constructed basic codes for dispersion relation analysis. This work is partially supported by Grants-in-Aid for Scientific Research (21H01083) and the Grant-in-Aid for Scientific Research on Innovative areas ``Unraveling the History of the Universe and Matter Evolution with Underground Physics" (19H05811) from the Ministry of Education, Culture, Sports, Science and Technology (MEXT), Japan. M.Z. is supported by the Japan Society for Promotion of Science (JSPS) Grant-in-Aid for JSPS Fellows (Grants No. 22J00440) from the Ministry of Education, Culture, Sports, Science, and Technology (MEXT) in Japan. S.Y. is supported by Institute for Advanced Theoretical and Experimental Physics, Waseda University, and the Waseda University Grant for Special Research Projects (project No. 2022C-140).
			\nocite{*}
			
			%\mz{Here is your reference bib. file.}
			\bibliography{apssamp.tex}% Produces the bibliography via BibTeX.

\providecommand{\noopsort}[1]{}\providecommand{\singleletter}[1]{#1}%
\begin{thebibliography}{38}%
\makeatletter
\providecommand \@ifxundefined [1]{%
 \@ifx{#1\undefined}
}%
\providecommand \@ifnum [1]{%
 \ifnum #1\expandafter \@firstoftwo
 \else \expandafter \@secondoftwo
 \fi
}%
\providecommand \@ifx [1]{%
 \ifx #1\expandafter \@firstoftwo
 \else \expandafter \@secondoftwo
 \fi
}%
\providecommand \natexlab [1]{#1}%
\providecommand \enquote  [1]{``#1''}%
\providecommand \bibnamefont  [1]{#1}%
\providecommand \bibfnamefont [1]{#1}%
\providecommand \citenamefont [1]{#1}%
\providecommand \href@noop [0]{\@secondoftwo}%
\providecommand \href [0]{\begingroup \@sanitize@url \@href}%
\providecommand \@href[1]{\@@startlink{#1}\@@href}%
\providecommand \@@href[1]{\endgroup#1\@@endlink}%
\providecommand \@sanitize@url [0]{\catcode `\\12\catcode `\$12\catcode
  `\&12\catcode `\#12\catcode `\^12\catcode `\_12\catcode `\%12\relax}%
\providecommand \@@startlink[1]{}%
\providecommand \@@endlink[0]{}%
\providecommand \url  [0]{\begingroup\@sanitize@url \@url }%
\providecommand \@url [1]{\endgroup\@href {#1}{\urlprefix }}%
\providecommand \urlprefix  [0]{URL }%
\providecommand \Eprint [0]{\href }%
\providecommand \doibase [0]{https://doi.org/}%
\providecommand \selectlanguage [0]{\@gobble}%
\providecommand \bibinfo  [0]{\@secondoftwo}%
\providecommand \bibfield  [0]{\@secondoftwo}%
\providecommand \translation [1]{[#1]}%
\providecommand \BibitemOpen [0]{}%
\providecommand \bibitemStop [0]{}%
\providecommand \bibitemNoStop [0]{.\EOS\space}%
\providecommand \EOS [0]{\spacefactor3000\relax}%
\providecommand \BibitemShut  [1]{\csname bibitem#1\endcsname}%
\let\auto@bib@innerbib\@empty
%</preamble>
\bibitem [{\citenamefont {Pantaleone}(1992)}]{PhysRevD.46.510}%
  \BibitemOpen
  \bibfield  {author} {\bibinfo {author} {\bibfnamefont {J.}~\bibnamefont
  {Pantaleone}},\ }\bibfield  {title} {\bibinfo {title} {Dirac neutrinos in
  dense matter},\ }\href {https://doi.org/10.1103/PhysRevD.46.510} {\bibfield
  {journal} {\bibinfo  {journal} {Phys. Rev. D}\ }\textbf {\bibinfo {volume}
  {46}},\ \bibinfo {pages} {510} (\bibinfo {year} {1992})}\BibitemShut
  {NoStop}%
\bibitem [{\citenamefont {Sigl}\ and\ \citenamefont
  {Raffelt}(1993)}]{Sigl1993GeneralKD}%
  \BibitemOpen
  \bibfield  {author} {\bibinfo {author} {\bibfnamefont {G.}~\bibnamefont
  {Sigl}}\ and\ \bibinfo {author} {\bibfnamefont {G.}~\bibnamefont {Raffelt}},\
  }\bibfield  {title} {\bibinfo {title} {General kinetic description of
  relativistic mixed neutrinos},\ }\href@noop {} {\bibfield  {journal}
  {\bibinfo  {journal} {Nuclear Physics}\ }\textbf {\bibinfo {volume} {406}},\
  \bibinfo {pages} {423} (\bibinfo {year} {1993})}\BibitemShut {NoStop}%
\bibitem [{\citenamefont {Duan}\ \emph {et~al.}(2010)\citenamefont {Duan},
  \citenamefont {Fuller},\ and\ \citenamefont
  {Qian}}]{doi:10.1146/annurev.nucl.012809.104524}%
  \BibitemOpen
  \bibfield  {author} {\bibinfo {author} {\bibfnamefont {H.}~\bibnamefont
  {Duan}}, \bibinfo {author} {\bibfnamefont {G.~M.}\ \bibnamefont {Fuller}},\
  and\ \bibinfo {author} {\bibfnamefont {Y.-Z.}\ \bibnamefont {Qian}},\
  }\bibfield  {title} {\bibinfo {title} {Collective neutrino oscillations},\
  }\href {https://doi.org/10.1146/annurev.nucl.012809.104524} {\bibfield
  {journal} {\bibinfo  {journal} {Annual Review of Nuclear and Particle
  Science}\ }\textbf {\bibinfo {volume} {60}},\ \bibinfo {pages} {569}
  (\bibinfo {year} {2010})},\ \Eprint
  {https://arxiv.org/abs/https://doi.org/10.1146/annurev.nucl.012809.104524}
  {https://doi.org/10.1146/annurev.nucl.012809.104524} \BibitemShut {NoStop}%
\bibitem [{\citenamefont {Chakraborty}\ \emph {et~al.}(2016)\citenamefont
  {Chakraborty}, \citenamefont {Hansen}, \citenamefont {Izaguirre},\ and\
  \citenamefont {Raffelt}}]{CHAKRABORTY2016366}%
  \BibitemOpen
  \bibfield  {author} {\bibinfo {author} {\bibfnamefont {S.}~\bibnamefont
  {Chakraborty}}, \bibinfo {author} {\bibfnamefont {R.}~\bibnamefont {Hansen}},
  \bibinfo {author} {\bibfnamefont {I.}~\bibnamefont {Izaguirre}},\ and\
  \bibinfo {author} {\bibfnamefont {G.}~\bibnamefont {Raffelt}},\ }\bibfield
  {title} {\bibinfo {title} {Collective neutrino flavor conversion: Recent
  developments},\ }\href
  {https://doi.org/https://doi.org/10.1016/j.nuclphysb.2016.02.012} {\bibfield
  {journal} {\bibinfo  {journal} {Nuclear Physics B}\ }\textbf {\bibinfo
  {volume} {908}},\ \bibinfo {pages} {366} (\bibinfo {year} {2016})},\ \bibinfo
  {note} {neutrino Oscillations: Celebrating the Nobel Prize in Physics
  2015}\BibitemShut {NoStop}%
\bibitem [{\citenamefont {Tamborra}\ and\ \citenamefont
  {Shalgar}(2021)}]{doi:10.1146/annurev-nucl-102920-050505}%
  \BibitemOpen
  \bibfield  {author} {\bibinfo {author} {\bibfnamefont {I.}~\bibnamefont
  {Tamborra}}\ and\ \bibinfo {author} {\bibfnamefont {S.}~\bibnamefont
  {Shalgar}},\ }\bibfield  {title} {\bibinfo {title} {New developments in
  flavor evolution of a dense neutrino gas},\ }\href
  {https://doi.org/10.1146/annurev-nucl-102920-050505} {\bibfield  {journal}
  {\bibinfo  {journal} {Annual Review of Nuclear and Particle Science}\
  }\textbf {\bibinfo {volume} {71}},\ \bibinfo {pages} {165} (\bibinfo {year}
  {2021})},\ \Eprint
  {https://arxiv.org/abs/https://doi.org/10.1146/annurev-nucl-102920-050505}
  {https://doi.org/10.1146/annurev-nucl-102920-050505} \BibitemShut {NoStop}%
\bibitem [{\citenamefont {Richers}\ and\ \citenamefont
  {Sen}(2022)}]{https://doi.org/10.48550/arxiv.2207.03561}%
  \BibitemOpen
  \bibfield  {author} {\bibinfo {author} {\bibfnamefont {S.}~\bibnamefont
  {Richers}}\ and\ \bibinfo {author} {\bibfnamefont {M.}~\bibnamefont {Sen}},\
  }\href {https://doi.org/10.48550/ARXIV.2207.03561} {\bibinfo {title} {Fast
  flavor transformations}} (\bibinfo {year} {2022})\BibitemShut {NoStop}%
\bibitem [{\citenamefont {Morinaga}(2022)}]{PhysRevD.105.L101301}%
  \BibitemOpen
  \bibfield  {author} {\bibinfo {author} {\bibfnamefont {T.}~\bibnamefont
  {Morinaga}},\ }\bibfield  {title} {\bibinfo {title} {Fast neutrino flavor
  instability and neutrino flavor lepton number crossings},\ }\href
  {https://doi.org/10.1103/PhysRevD.105.L101301} {\bibfield  {journal}
  {\bibinfo  {journal} {Phys. Rev. D}\ }\textbf {\bibinfo {volume} {105}},\
  \bibinfo {pages} {L101301} (\bibinfo {year} {2022})}\BibitemShut {NoStop}%
\bibitem [{\citenamefont {Abbar}\ \emph {et~al.}(2019)\citenamefont {Abbar},
  \citenamefont {Duan}, \citenamefont {Sumiyoshi}, \citenamefont {Takiwaki},\
  and\ \citenamefont {Volpe}}]{PhysRevD.100.043004}%
  \BibitemOpen
  \bibfield  {author} {\bibinfo {author} {\bibfnamefont {S.}~\bibnamefont
  {Abbar}}, \bibinfo {author} {\bibfnamefont {H.}~\bibnamefont {Duan}},
  \bibinfo {author} {\bibfnamefont {K.}~\bibnamefont {Sumiyoshi}}, \bibinfo
  {author} {\bibfnamefont {T.}~\bibnamefont {Takiwaki}},\ and\ \bibinfo
  {author} {\bibfnamefont {M.~C.}\ \bibnamefont {Volpe}},\ }\bibfield  {title}
  {\bibinfo {title} {On the occurrence of fast neutrino flavor conversions in
  multidimensional supernova models},\ }\href
  {https://doi.org/10.1103/PhysRevD.100.043004} {\bibfield  {journal} {\bibinfo
   {journal} {Phys. Rev. D}\ }\textbf {\bibinfo {volume} {100}},\ \bibinfo
  {pages} {043004} (\bibinfo {year} {2019})}\BibitemShut {NoStop}%
\bibitem [{\citenamefont {Nagakura}\ \emph {et~al.}(2019)\citenamefont
  {Nagakura}, \citenamefont {Morinaga}, \citenamefont {Kato},\ and\
  \citenamefont {Yamada}}]{Nagakura_2019}%
  \BibitemOpen
  \bibfield  {author} {\bibinfo {author} {\bibfnamefont {H.}~\bibnamefont
  {Nagakura}}, \bibinfo {author} {\bibfnamefont {T.}~\bibnamefont {Morinaga}},
  \bibinfo {author} {\bibfnamefont {C.}~\bibnamefont {Kato}},\ and\ \bibinfo
  {author} {\bibfnamefont {S.}~\bibnamefont {Yamada}},\ }\bibfield  {title}
  {\bibinfo {title} {Fast-pairwise collective neutrino oscillations associated
  with asymmetric neutrino emissions in core-collapse supernovae},\ }\href
  {https://doi.org/10.3847/1538-4357/ab4cf2} {\bibfield  {journal} {\bibinfo
  {journal} {The Astrophysical Journal}\ }\textbf {\bibinfo {volume} {886}},\
  \bibinfo {pages} {139} (\bibinfo {year} {2019})}\BibitemShut {NoStop}%
\bibitem [{\citenamefont {Delfan~Azari}\ \emph {et~al.}(2020)\citenamefont
  {Delfan~Azari}, \citenamefont {Yamada}, \citenamefont {Morinaga},
  \citenamefont {Nagakura}, \citenamefont {Furusawa}, \citenamefont {Harada},
  \citenamefont {Okawa}, \citenamefont {Iwakami},\ and\ \citenamefont
  {Sumiyoshi}}]{PhysRevD.101.023018}%
  \BibitemOpen
  \bibfield  {author} {\bibinfo {author} {\bibfnamefont {M.}~\bibnamefont
  {Delfan~Azari}}, \bibinfo {author} {\bibfnamefont {S.}~\bibnamefont
  {Yamada}}, \bibinfo {author} {\bibfnamefont {T.}~\bibnamefont {Morinaga}},
  \bibinfo {author} {\bibfnamefont {H.}~\bibnamefont {Nagakura}}, \bibinfo
  {author} {\bibfnamefont {S.}~\bibnamefont {Furusawa}}, \bibinfo {author}
  {\bibfnamefont {A.}~\bibnamefont {Harada}}, \bibinfo {author} {\bibfnamefont
  {H.}~\bibnamefont {Okawa}}, \bibinfo {author} {\bibfnamefont
  {W.}~\bibnamefont {Iwakami}},\ and\ \bibinfo {author} {\bibfnamefont
  {K.}~\bibnamefont {Sumiyoshi}},\ }\bibfield  {title} {\bibinfo {title} {Fast
  collective neutrino oscillations inside the neutrino sphere in core-collapse
  supernovae},\ }\href {https://doi.org/10.1103/PhysRevD.101.023018} {\bibfield
   {journal} {\bibinfo  {journal} {Phys. Rev. D}\ }\textbf {\bibinfo {volume}
  {101}},\ \bibinfo {pages} {023018} (\bibinfo {year} {2020})}\BibitemShut
  {NoStop}%
\bibitem [{\citenamefont {Abbar}\ \emph {et~al.}(2020)\citenamefont {Abbar},
  \citenamefont {Duan}, \citenamefont {Sumiyoshi}, \citenamefont {Takiwaki},\
  and\ \citenamefont {Volpe}}]{PhysRevD.101.043016}%
  \BibitemOpen
  \bibfield  {author} {\bibinfo {author} {\bibfnamefont {S.}~\bibnamefont
  {Abbar}}, \bibinfo {author} {\bibfnamefont {H.}~\bibnamefont {Duan}},
  \bibinfo {author} {\bibfnamefont {K.}~\bibnamefont {Sumiyoshi}}, \bibinfo
  {author} {\bibfnamefont {T.}~\bibnamefont {Takiwaki}},\ and\ \bibinfo
  {author} {\bibfnamefont {M.~C.}\ \bibnamefont {Volpe}},\ }\bibfield  {title}
  {\bibinfo {title} {Fast neutrino flavor conversion modes in multidimensional
  core-collapse supernova models: The role of the asymmetric neutrino
  distributions},\ }\href {https://doi.org/10.1103/PhysRevD.101.043016}
  {\bibfield  {journal} {\bibinfo  {journal} {Phys. Rev. D}\ }\textbf {\bibinfo
  {volume} {101}},\ \bibinfo {pages} {043016} (\bibinfo {year}
  {2020})}\BibitemShut {NoStop}%
\bibitem [{\citenamefont {Morinaga}\ \emph {et~al.}(2020)\citenamefont
  {Morinaga}, \citenamefont {Nagakura}, \citenamefont {Kato},\ and\
  \citenamefont {Yamada}}]{PhysRevResearch.2.012046}%
  \BibitemOpen
  \bibfield  {author} {\bibinfo {author} {\bibfnamefont {T.}~\bibnamefont
  {Morinaga}}, \bibinfo {author} {\bibfnamefont {H.}~\bibnamefont {Nagakura}},
  \bibinfo {author} {\bibfnamefont {C.}~\bibnamefont {Kato}},\ and\ \bibinfo
  {author} {\bibfnamefont {S.}~\bibnamefont {Yamada}},\ }\bibfield  {title}
  {\bibinfo {title} {Fast neutrino-flavor conversion in the preshock region of
  core-collapse supernovae},\ }\href
  {https://doi.org/10.1103/PhysRevResearch.2.012046} {\bibfield  {journal}
  {\bibinfo  {journal} {Phys. Rev. Res.}\ }\textbf {\bibinfo {volume} {2}},\
  \bibinfo {pages} {012046} (\bibinfo {year} {2020})}\BibitemShut {NoStop}%
\bibitem [{\citenamefont {Glas}\ \emph {et~al.}(2020)\citenamefont {Glas},
  \citenamefont {Janka}, \citenamefont {Capozzi}, \citenamefont {Sen},
  \citenamefont {Dasgupta}, \citenamefont {Mirizzi},\ and\ \citenamefont
  {Sigl}}]{PhysRevD.101.063001}%
  \BibitemOpen
  \bibfield  {author} {\bibinfo {author} {\bibfnamefont {R.}~\bibnamefont
  {Glas}}, \bibinfo {author} {\bibfnamefont {H.-T.}\ \bibnamefont {Janka}},
  \bibinfo {author} {\bibfnamefont {F.}~\bibnamefont {Capozzi}}, \bibinfo
  {author} {\bibfnamefont {M.}~\bibnamefont {Sen}}, \bibinfo {author}
  {\bibfnamefont {B.}~\bibnamefont {Dasgupta}}, \bibinfo {author}
  {\bibfnamefont {A.}~\bibnamefont {Mirizzi}},\ and\ \bibinfo {author}
  {\bibfnamefont {G.}~\bibnamefont {Sigl}},\ }\bibfield  {title} {\bibinfo
  {title} {Fast neutrino flavor instability in the neutron-star convection
  layer of three-dimensional supernova models},\ }\href
  {https://doi.org/10.1103/PhysRevD.101.063001} {\bibfield  {journal} {\bibinfo
   {journal} {Phys. Rev. D}\ }\textbf {\bibinfo {volume} {101}},\ \bibinfo
  {pages} {063001} (\bibinfo {year} {2020})}\BibitemShut {NoStop}%
\bibitem [{\citenamefont {Abbar}(2020)}]{Abbar_2020}%
  \BibitemOpen
  \bibfield  {author} {\bibinfo {author} {\bibfnamefont {S.}~\bibnamefont
  {Abbar}},\ }\bibfield  {title} {\bibinfo {title} {Searching for fast neutrino
  flavor conversion modes in core-collapse supernova simulations},\ }\href
  {https://doi.org/10.1088/1475-7516/2020/05/027} {\bibfield  {journal}
  {\bibinfo  {journal} {Journal of Cosmology and Astroparticle Physics}\
  }\textbf {\bibinfo {volume} {2020}}\bibinfo  {number} { (05)},\ \bibinfo
  {pages} {027}}\BibitemShut {NoStop}%
\bibitem [{\citenamefont {Capozzi}\ \emph {et~al.}(2021)\citenamefont
  {Capozzi}, \citenamefont {Abbar}, \citenamefont {Bollig},\ and\ \citenamefont
  {Janka}}]{PhysRevD.103.063013}%
  \BibitemOpen
\bibfield  {number} {  }\bibfield  {author} {\bibinfo {author} {\bibfnamefont
  {F.}~\bibnamefont {Capozzi}}, \bibinfo {author} {\bibfnamefont
  {S.}~\bibnamefont {Abbar}}, \bibinfo {author} {\bibfnamefont
  {R.}~\bibnamefont {Bollig}},\ and\ \bibinfo {author} {\bibfnamefont {H.-T.}\
  \bibnamefont {Janka}},\ }\bibfield  {title} {\bibinfo {title} {Fast neutrino
  flavor conversions in one-dimensional core-collapse supernova models with and
  without muon creation},\ }\href {https://doi.org/10.1103/PhysRevD.103.063013}
  {\bibfield  {journal} {\bibinfo  {journal} {Phys. Rev. D}\ }\textbf {\bibinfo
  {volume} {103}},\ \bibinfo {pages} {063013} (\bibinfo {year}
  {2021})}\BibitemShut {NoStop}%
\bibitem [{\citenamefont {Nagakura}\ \emph {et~al.}(2021)\citenamefont
  {Nagakura}, \citenamefont {Burrows}, \citenamefont {Johns},\ and\
  \citenamefont {Fuller}}]{PhysRevD.104.083025}%
  \BibitemOpen
  \bibfield  {author} {\bibinfo {author} {\bibfnamefont {H.}~\bibnamefont
  {Nagakura}}, \bibinfo {author} {\bibfnamefont {A.}~\bibnamefont {Burrows}},
  \bibinfo {author} {\bibfnamefont {L.}~\bibnamefont {Johns}},\ and\ \bibinfo
  {author} {\bibfnamefont {G.~M.}\ \bibnamefont {Fuller}},\ }\bibfield  {title}
  {\bibinfo {title} {Where, when, and why: Occurrence of fast-pairwise
  collective neutrino oscillation in three-dimensional core-collapse supernova
  models},\ }\href {https://doi.org/10.1103/PhysRevD.104.083025} {\bibfield
  {journal} {\bibinfo  {journal} {Phys. Rev. D}\ }\textbf {\bibinfo {volume}
  {104}},\ \bibinfo {pages} {083025} (\bibinfo {year} {2021})}\BibitemShut
  {NoStop}%
\bibitem [{\citenamefont {Harada}\ and\ \citenamefont
  {Nagakura}(2022)}]{Harada_2022}%
  \BibitemOpen
  \bibfield  {author} {\bibinfo {author} {\bibfnamefont {A.}~\bibnamefont
  {Harada}}\ and\ \bibinfo {author} {\bibfnamefont {H.}~\bibnamefont
  {Nagakura}},\ }\bibfield  {title} {\bibinfo {title} {Prospects of fast flavor
  neutrino conversion in rotating core-collapse supernovae},\ }\href
  {https://doi.org/10.3847/1538-4357/ac38a0} {\bibfield  {journal} {\bibinfo
  {journal} {The Astrophysical Journal}\ }\textbf {\bibinfo {volume} {924}},\
  \bibinfo {pages} {109} (\bibinfo {year} {2022})}\BibitemShut {NoStop}%
\bibitem [{\citenamefont {Akaho}\ \emph {et~al.}(2022)\citenamefont {Akaho},
  \citenamefont {Harada}, \citenamefont {Nagakura}, \citenamefont {Iwakami},
  \citenamefont {Okawa}, \citenamefont {Furusawa}, \citenamefont {Matsufuru},
  \citenamefont {Sumiyoshi},\ and\ \citenamefont
  {Yamada}}]{https://doi.org/10.48550/arxiv.2206.01673}%
  \BibitemOpen
  \bibfield  {author} {\bibinfo {author} {\bibfnamefont {R.}~\bibnamefont
  {Akaho}}, \bibinfo {author} {\bibfnamefont {A.}~\bibnamefont {Harada}},
  \bibinfo {author} {\bibfnamefont {H.}~\bibnamefont {Nagakura}}, \bibinfo
  {author} {\bibfnamefont {W.}~\bibnamefont {Iwakami}}, \bibinfo {author}
  {\bibfnamefont {H.}~\bibnamefont {Okawa}}, \bibinfo {author} {\bibfnamefont
  {S.}~\bibnamefont {Furusawa}}, \bibinfo {author} {\bibfnamefont
  {H.}~\bibnamefont {Matsufuru}}, \bibinfo {author} {\bibfnamefont
  {K.}~\bibnamefont {Sumiyoshi}},\ and\ \bibinfo {author} {\bibfnamefont
  {S.}~\bibnamefont {Yamada}},\ }\href
  {https://doi.org/10.48550/ARXIV.2206.01673} {\bibinfo {title} {Protoneutron
  star convection simulated with a new general relativistic boltzmann neutrino
  radiation-hydrodynamics code}} (\bibinfo {year} {2022})\BibitemShut {NoStop}%
\bibitem [{\citenamefont {Martin}\ \emph {et~al.}(2021)\citenamefont {Martin},
  \citenamefont {Carlson}, \citenamefont {Cirigliano},\ and\ \citenamefont
  {Duan}}]{PhysRevD.103.063001}%
  \BibitemOpen
  \bibfield  {author} {\bibinfo {author} {\bibfnamefont {J.~D.}\ \bibnamefont
  {Martin}}, \bibinfo {author} {\bibfnamefont {J.}~\bibnamefont {Carlson}},
  \bibinfo {author} {\bibfnamefont {V.}~\bibnamefont {Cirigliano}},\ and\
  \bibinfo {author} {\bibfnamefont {H.}~\bibnamefont {Duan}},\ }\bibfield
  {title} {\bibinfo {title} {Fast flavor oscillations in dense neutrino media
  with collisions},\ }\href {https://doi.org/10.1103/PhysRevD.103.063001}
  {\bibfield  {journal} {\bibinfo  {journal} {Phys. Rev. D}\ }\textbf {\bibinfo
  {volume} {103}},\ \bibinfo {pages} {063001} (\bibinfo {year}
  {2021})}\BibitemShut {NoStop}%
\bibitem [{\citenamefont {Sigl}(2022)}]{PhysRevD.105.043005}%
  \BibitemOpen
  \bibfield  {author} {\bibinfo {author} {\bibfnamefont {G.}~\bibnamefont
  {Sigl}},\ }\bibfield  {title} {\bibinfo {title} {Simulations of fast neutrino
  flavor conversions with interactions in inhomogeneous media},\ }\href
  {https://doi.org/10.1103/PhysRevD.105.043005} {\bibfield  {journal} {\bibinfo
   {journal} {Phys. Rev. D}\ }\textbf {\bibinfo {volume} {105}},\ \bibinfo
  {pages} {043005} (\bibinfo {year} {2022})}\BibitemShut {NoStop}%
\bibitem [{\citenamefont {Johns}\ and\ \citenamefont
  {Nagakura}(2022)}]{PhysRevD.106.043031}%
  \BibitemOpen
  \bibfield  {author} {\bibinfo {author} {\bibfnamefont {L.}~\bibnamefont
  {Johns}}\ and\ \bibinfo {author} {\bibfnamefont {H.}~\bibnamefont
  {Nagakura}},\ }\bibfield  {title} {\bibinfo {title} {Self-consistency in
  models of neutrino scattering and fast flavor conversion},\ }\href
  {https://doi.org/10.1103/PhysRevD.106.043031} {\bibfield  {journal} {\bibinfo
   {journal} {Phys. Rev. D}\ }\textbf {\bibinfo {volume} {106}},\ \bibinfo
  {pages} {043031} (\bibinfo {year} {2022})}\BibitemShut {NoStop}%
\bibitem [{\citenamefont {Shalgar}\ and\ \citenamefont
  {Tamborra}(2021)}]{PhysRevD.103.063002}%
  \BibitemOpen
  \bibfield  {author} {\bibinfo {author} {\bibfnamefont {S.}~\bibnamefont
  {Shalgar}}\ and\ \bibinfo {author} {\bibfnamefont {I.}~\bibnamefont
  {Tamborra}},\ }\bibfield  {title} {\bibinfo {title} {Change of direction in
  pairwise neutrino conversion physics: The effect of collisions},\ }\href
  {https://doi.org/10.1103/PhysRevD.103.063002} {\bibfield  {journal} {\bibinfo
   {journal} {Phys. Rev. D}\ }\textbf {\bibinfo {volume} {103}},\ \bibinfo
  {pages} {063002} (\bibinfo {year} {2021})}\BibitemShut {NoStop}%
\bibitem [{\citenamefont {Kato}\ \emph {et~al.}(2021)\citenamefont {Kato},
  \citenamefont {Nagakura},\ and\ \citenamefont {Morinaga}}]{Kato_2021}%
  \BibitemOpen
  \bibfield  {author} {\bibinfo {author} {\bibfnamefont {C.}~\bibnamefont
  {Kato}}, \bibinfo {author} {\bibfnamefont {H.}~\bibnamefont {Nagakura}},\
  and\ \bibinfo {author} {\bibfnamefont {T.}~\bibnamefont {Morinaga}},\
  }\bibfield  {title} {\bibinfo {title} {Neutrino transport with the monte
  carlo method. ii. quantum kinetic equations},\ }\href
  {https://doi.org/10.3847/1538-4365/ac2aa4} {\bibfield  {journal} {\bibinfo
  {journal} {The Astrophysical Journal Supplement Series}\ }\textbf {\bibinfo
  {volume} {257}},\ \bibinfo {pages} {55} (\bibinfo {year} {2021})}\BibitemShut
  {NoStop}%
\bibitem [{\citenamefont {Sasaki}\ and\ \citenamefont
  {Takiwaki}(2022)}]{10.1093/ptep/ptac082}%
  \BibitemOpen
  \bibfield  {author} {\bibinfo {author} {\bibfnamefont {H.}~\bibnamefont
  {Sasaki}}\ and\ \bibinfo {author} {\bibfnamefont {T.}~\bibnamefont
  {Takiwaki}},\ }\bibfield  {title} {\bibinfo {title} {{A detailed analysis of
  the dynamics of fast neutrino flavor conversions with scattering effects}},\
  }\bibfield  {journal} {\bibinfo  {journal} {Progress of Theoretical and
  Experimental Physics}\ }\textbf {\bibinfo {volume} {2022}},\ \href
  {https://doi.org/10.1093/ptep/ptac082} {10.1093/ptep/ptac082} (\bibinfo
  {year} {2022}),\ \bibinfo {note} {073E01},\ \Eprint
  {https://arxiv.org/abs/https://academic.oup.com/ptep/article-pdf/2022/7/073E01/44400755/ptac082.pdf}
  {https://academic.oup.com/ptep/article-pdf/2022/7/073E01/44400755/ptac082.pdf}
  \BibitemShut {NoStop}%
\bibitem [{\citenamefont {Hansen}\ \emph {et~al.}(2022)\citenamefont {Hansen},
  \citenamefont {Shalgar},\ and\ \citenamefont
  {Tamborra}}]{PhysRevD.105.123003}%
  \BibitemOpen
  \bibfield  {author} {\bibinfo {author} {\bibfnamefont {R.~S.~L.}\
  \bibnamefont {Hansen}}, \bibinfo {author} {\bibfnamefont {S.}~\bibnamefont
  {Shalgar}},\ and\ \bibinfo {author} {\bibfnamefont {I.}~\bibnamefont
  {Tamborra}},\ }\bibfield  {title} {\bibinfo {title} {Enhancement or damping
  of fast neutrino flavor conversions due to collisions},\ }\href
  {https://doi.org/10.1103/PhysRevD.105.123003} {\bibfield  {journal} {\bibinfo
   {journal} {Phys. Rev. D}\ }\textbf {\bibinfo {volume} {105}},\ \bibinfo
  {pages} {123003} (\bibinfo {year} {2022})}\BibitemShut {NoStop}%
\bibitem [{\citenamefont {Kato}\ and\ \citenamefont
  {Nagakura}(2022)}]{Kato_2022}%
  \BibitemOpen
  \bibfield  {author} {\bibinfo {author} {\bibfnamefont {C.}~\bibnamefont
  {Kato}}\ and\ \bibinfo {author} {\bibfnamefont {H.}~\bibnamefont
  {Nagakura}},\ }\bibfield  {title} {\bibinfo {title} {Effects of
  energy-dependent scatterings on fast neutrino flavor conversions},\
  }\bibfield  {journal} {\bibinfo  {journal} {Physical Review D}\ }\textbf
  {\bibinfo {volume} {106}},\ \href
  {https://doi.org/10.1103/physrevd.106.123013} {10.1103/physrevd.106.123013}
  (\bibinfo {year} {2022})\BibitemShut {NoStop}%
\bibitem [{\citenamefont {Padilla-Gay}\ \emph {et~al.}(2022)\citenamefont
  {Padilla-Gay}, \citenamefont {Tamborra},\ and\ \citenamefont
  {Raffelt}}]{PhysRevD.106.103031}%
  \BibitemOpen
  \bibfield  {author} {\bibinfo {author} {\bibfnamefont {I.}~\bibnamefont
  {Padilla-Gay}}, \bibinfo {author} {\bibfnamefont {I.}~\bibnamefont
  {Tamborra}},\ and\ \bibinfo {author} {\bibfnamefont {G.~G.}\ \bibnamefont
  {Raffelt}},\ }\bibfield  {title} {\bibinfo {title} {Neutrino fast flavor
  pendulum. ii. collisional damping},\ }\href
  {https://doi.org/10.1103/PhysRevD.106.103031} {\bibfield  {journal} {\bibinfo
   {journal} {Phys. Rev. D}\ }\textbf {\bibinfo {volume} {106}},\ \bibinfo
  {pages} {103031} (\bibinfo {year} {2022})}\BibitemShut {NoStop}%
\bibitem [{\citenamefont {Capozzi}\ \emph {et~al.}(2019)\citenamefont
  {Capozzi}, \citenamefont {Dasgupta}, \citenamefont {Mirizzi}, \citenamefont
  {Sen},\ and\ \citenamefont {Sigl}}]{PhysRevLett.122.091101}%
  \BibitemOpen
  \bibfield  {author} {\bibinfo {author} {\bibfnamefont {F.}~\bibnamefont
  {Capozzi}}, \bibinfo {author} {\bibfnamefont {B.}~\bibnamefont {Dasgupta}},
  \bibinfo {author} {\bibfnamefont {A.}~\bibnamefont {Mirizzi}}, \bibinfo
  {author} {\bibfnamefont {M.}~\bibnamefont {Sen}},\ and\ \bibinfo {author}
  {\bibfnamefont {G.}~\bibnamefont {Sigl}},\ }\bibfield  {title} {\bibinfo
  {title} {Collisional triggering of fast flavor conversions of supernova
  neutrinos},\ }\href {https://doi.org/10.1103/PhysRevLett.122.091101}
  {\bibfield  {journal} {\bibinfo  {journal} {Phys. Rev. Lett.}\ }\textbf
  {\bibinfo {volume} {122}},\ \bibinfo {pages} {091101} (\bibinfo {year}
  {2019})}\BibitemShut {NoStop}%
\bibitem [{\citenamefont {Johns}(2021)}]{luke2019}%
  \BibitemOpen
  \bibfield  {author} {\bibinfo {author} {\bibfnamefont {L.}~\bibnamefont
  {Johns}},\ }\href {https://doi.org/10.48550/ARXIV.2104.11369} {\bibinfo
  {title} {Collisional flavor instabilities of supernova neutrinos}} (\bibinfo
  {year} {2021})\BibitemShut {NoStop}%
\bibitem [{\citenamefont {Johns}\ and\ \citenamefont
  {Xiong}(2022)}]{PhysRevD.106.103029}%
  \BibitemOpen
  \bibfield  {author} {\bibinfo {author} {\bibfnamefont {L.}~\bibnamefont
  {Johns}}\ and\ \bibinfo {author} {\bibfnamefont {Z.}~\bibnamefont {Xiong}},\
  }\bibfield  {title} {\bibinfo {title} {Collisional instabilities of neutrinos
  and their interplay with fast flavor conversion in compact objects},\ }\href
  {https://doi.org/10.1103/PhysRevD.106.103029} {\bibfield  {journal} {\bibinfo
   {journal} {Phys. Rev. D}\ }\textbf {\bibinfo {volume} {106}},\ \bibinfo
  {pages} {103029} (\bibinfo {year} {2022})}\BibitemShut {NoStop}%
\bibitem [{\citenamefont {Lin}\ and\ \citenamefont {Duan}(2022)}]{duan}%
  \BibitemOpen
  \bibfield  {author} {\bibinfo {author} {\bibfnamefont {Y.-C.}\ \bibnamefont
  {Lin}}\ and\ \bibinfo {author} {\bibfnamefont {H.}~\bibnamefont {Duan}},\
  }\href {https://doi.org/10.48550/ARXIV.2210.09218} {\bibinfo {title}
  {Collision-induced flavor instability in dense neutrino gases with
  energy-dependent scattering}} (\bibinfo {year} {2022})\BibitemShut {NoStop}%
\bibitem [{\citenamefont {Xiong}\ \emph {et~al.}(2023)\citenamefont {Xiong},
  \citenamefont {Wu}, \citenamefont {Mart\'{\i}nez-Pinedo}, \citenamefont
  {Fischer}, \citenamefont {George}, \citenamefont {Lin},\ and\ \citenamefont
  {Johns}}]{PhysRevD.107.083016}%
  \BibitemOpen
  \bibfield  {author} {\bibinfo {author} {\bibfnamefont {Z.}~\bibnamefont
  {Xiong}}, \bibinfo {author} {\bibfnamefont {M.-R.}\ \bibnamefont {Wu}},
  \bibinfo {author} {\bibfnamefont {G.}~\bibnamefont {Mart\'{\i}nez-Pinedo}},
  \bibinfo {author} {\bibfnamefont {T.}~\bibnamefont {Fischer}}, \bibinfo
  {author} {\bibfnamefont {M.}~\bibnamefont {George}}, \bibinfo {author}
  {\bibfnamefont {C.-Y.}\ \bibnamefont {Lin}},\ and\ \bibinfo {author}
  {\bibfnamefont {L.}~\bibnamefont {Johns}},\ }\bibfield  {title} {\bibinfo
  {title} {Evolution of collisional neutrino flavor instabilities in
  spherically symmetric supernova models},\ }\href
  {https://doi.org/10.1103/PhysRevD.107.083016} {\bibfield  {journal} {\bibinfo
   {journal} {Phys. Rev. D}\ }\textbf {\bibinfo {volume} {107}},\ \bibinfo
  {pages} {083016} (\bibinfo {year} {2023})}\BibitemShut {NoStop}%
\bibitem [{\citenamefont {Xiong}\ \emph {et~al.}(2022)\citenamefont {Xiong},
  \citenamefont {Johns}, \citenamefont {Wu},\ and\ \citenamefont
  {Duan}}]{2022}%
  \BibitemOpen
  \bibfield  {author} {\bibinfo {author} {\bibfnamefont {Z.}~\bibnamefont
  {Xiong}}, \bibinfo {author} {\bibfnamefont {L.}~\bibnamefont {Johns}},
  \bibinfo {author} {\bibfnamefont {M.-R.}\ \bibnamefont {Wu}},\ and\ \bibinfo
  {author} {\bibfnamefont {H.}~\bibnamefont {Duan}},\ }\href
  {https://doi.org/10.48550/ARXIV.2212.03750} {\bibinfo {title} {Collisional
  flavor instability in dense neutrino gases}} (\bibinfo {year}
  {2022})\BibitemShut {NoStop}%
\bibitem [{\citenamefont {Airen}\ \emph {et~al.}(2018)\citenamefont {Airen},
  \citenamefont {Capozzi}, \citenamefont {Chakraborty}, \citenamefont
  {Dasgupta}, \citenamefont {Raffelt},\ and\ \citenamefont
  {Stirner}}]{Airen_2018}%
  \BibitemOpen
  \bibfield  {author} {\bibinfo {author} {\bibfnamefont {S.}~\bibnamefont
  {Airen}}, \bibinfo {author} {\bibfnamefont {F.}~\bibnamefont {Capozzi}},
  \bibinfo {author} {\bibfnamefont {S.}~\bibnamefont {Chakraborty}}, \bibinfo
  {author} {\bibfnamefont {B.}~\bibnamefont {Dasgupta}}, \bibinfo {author}
  {\bibfnamefont {G.}~\bibnamefont {Raffelt}},\ and\ \bibinfo {author}
  {\bibfnamefont {T.}~\bibnamefont {Stirner}},\ }\bibfield  {title} {\bibinfo
  {title} {Normal-mode analysis for collective neutrino oscillations},\ }\href
  {https://doi.org/10.1088/1475-7516/2018/12/019} {\bibfield  {journal}
  {\bibinfo  {journal} {Journal of Cosmology and Astroparticle Physics}\
  }\textbf {\bibinfo {volume} {2018}}\bibinfo  {number} { (12)},\ \bibinfo
  {pages} {019}}\BibitemShut {NoStop}%
\bibitem [{\citenamefont {Izaguirre}\ \emph {et~al.}(2017)\citenamefont
  {Izaguirre}, \citenamefont {Raffelt},\ and\ \citenamefont
  {Tamborra}}]{PhysRevLett.118.021101}%
  \BibitemOpen
\bibfield  {number} {  }\bibfield  {author} {\bibinfo {author} {\bibfnamefont
  {I.}~\bibnamefont {Izaguirre}}, \bibinfo {author} {\bibfnamefont
  {G.}~\bibnamefont {Raffelt}},\ and\ \bibinfo {author} {\bibfnamefont
  {I.}~\bibnamefont {Tamborra}},\ }\bibfield  {title} {\bibinfo {title} {Fast
  pairwise conversion of supernova neutrinos: A dispersion relation approach},\
  }\href {https://doi.org/10.1103/PhysRevLett.118.021101} {\bibfield  {journal}
  {\bibinfo  {journal} {Phys. Rev. Lett.}\ }\textbf {\bibinfo {volume} {118}},\
  \bibinfo {pages} {021101} (\bibinfo {year} {2017})}\BibitemShut {NoStop}%
\bibitem [{\citenamefont {{Zaizen}}\ and\ \citenamefont
  {{Nagakura}}(2022)}]{finalstate}%
  \BibitemOpen
  \bibfield  {author} {\bibinfo {author} {\bibfnamefont {M.}~\bibnamefont
  {{Zaizen}}}\ and\ \bibinfo {author} {\bibfnamefont {H.}~\bibnamefont
  {{Nagakura}}},\ }\bibfield  {title} {\bibinfo {title} {{Simple method for
  determining asymptotic states of fast neutrino-flavor conversion}},\
  }\href@noop {} {\bibfield  {journal} {\bibinfo  {journal} {arXiv e-prints}\
  ,\ \bibinfo {eid} {arXiv:2211.09343}} (\bibinfo {year} {2022})},\ \Eprint
  {https://arxiv.org/abs/2211.09343} {arXiv:2211.09343 [astro-ph.HE]}
  \BibitemShut {NoStop}%
\bibitem [{\citenamefont {Shalgar}\ and\ \citenamefont
  {Tamborra}(2019)}]{Shalgar_2019}%
  \BibitemOpen
  \bibfield  {author} {\bibinfo {author} {\bibfnamefont {S.}~\bibnamefont
  {Shalgar}}\ and\ \bibinfo {author} {\bibfnamefont {I.}~\bibnamefont
  {Tamborra}},\ }\bibfield  {title} {\bibinfo {title} {On the occurrence of
  crossings between the angular distributions of electron neutrinos and
  antineutrinos in the supernova core},\ }\href
  {https://doi.org/10.3847/1538-4357/ab38ba} {\bibfield  {journal} {\bibinfo
  {journal} {The Astrophysical Journal}\ }\textbf {\bibinfo {volume} {883}},\
  \bibinfo {pages} {80} (\bibinfo {year} {2019})}\BibitemShut {NoStop}%
\bibitem [{\citenamefont {Shalgar}\ and\ \citenamefont
  {Tamborra}(2022)}]{https://doi.org/10.48550/arxiv.2207.04058}%
  \BibitemOpen
  \bibfield  {author} {\bibinfo {author} {\bibfnamefont {S.}~\bibnamefont
  {Shalgar}}\ and\ \bibinfo {author} {\bibfnamefont {I.}~\bibnamefont
  {Tamborra}},\ }\href {https://doi.org/10.48550/ARXIV.2207.04058} {\bibinfo
  {title} {Neutrino flavor conversion, advection, and collisions: The full
  solution}} (\bibinfo {year} {2022})\BibitemShut {NoStop}%
\end{thebibliography}%
			
		\end{document}